\documentclass[12pt]{article}
\usepackage{amsmath,amssymb}
\usepackage{graphicx}
\usepackage{natbib}
\usepackage{bm, bbm}
\usepackage{subfig}
\usepackage{makecell}

\newcommand{\iid}{\stackrel{\rm iid}{\sim}}

\begin{document}

\title{Joint modelling of association networks and longitudinal biomarkers: an application to child obesity}

\author{Andrea Cremaschi $^{1,\ast}$, Maria De Iorio$^{1,2,3,4}$, Narasimhan Kothandaraman$^{1}$, \\ Fabian Yap$^{5}$, Mya Tway Tint$^{1}$, Johan Eriksson$^{1,2}$
	\\[2pt]
	\textit{$^{1}$Singapore Institute for Clinical Sciences, A*STAR, Singapore}\\
	\textit{$^{2}$Yong Loo Lin School of Medicine, National University of Singapore, Singapore}\\
	\textit{$^{3}$Division of Science, Yale-NUS College, Singapore}\\
	\textit{$^{4}$Department of Statistical Science, University College London, UK}\\
	\textit{$^{5}$Department of Paediatrics, KK Women’s and Children’s Hospital, Singapore}
	\\[2pt]
	{cremaschia@sics.a-star.edu.sg}}


\date{}

\maketitle


\bigskip
\begin{abstract}
	{The prevalence of chronic non-communicable diseases such as obesity has noticeably increased in the last decade. The study of these diseases in early life is of paramount importance in determining their course in adult life and in supporting clinical interventions. Recently, attention has been drawn on approaches that study the alteration of metabolic pathways in obese children. In this work, we propose a novel joint modelling approach for the analysis of growth biomarkers and metabolite concentrations, to unveil metabolic pathways related to child obesity. Within a Bayesian framework, we flexibly model the temporal evolution of growth trajectories and metabolic associations through the specification of a joint non-parametric random effect distribution which also allows for clustering of the subjects, thus identifying risk sub-groups. Growth profiles as well as patterns of metabolic associations determine the clustering structure. Inclusion of risk factors is straightforward through the specification of a regression term. We demonstrate the  proposed approach on data from the Growing Up in Singapore Towards healthy Outcomes (GUSTO) cohort study, based in Singapore. Posterior inference is obtained via a tailored MCMC algorithm, accommodating a nonparametric prior with mixed support. Our analysis has identified potential key pathways in obese children that allows for exploration of possible molecular mechanisms associated with child obesity.}
	{Dirichlet Process; Gaussian process; Graph-based clustering;  Graphical models; Longitudinal data;  Metabolomics} 
\end{abstract}

\section{Introduction}\label{sec:Intro}

Obesity is a major risk factor for chronic non-communicable diseases such as type-2 diabetes (T2D), metabolic syndrome and cardiovascular diseases. The prevalence of obesity has reached epidemic proportions worldwide and has tripled between 1975 and 2016. In particular in 2016, 39\% of adults were overweight and 13\% obese \citep{WHO_Obesity_and_overweight}. Prevalence of obesity in children has also escalated over the years, increasing from 4\% in 1975 to over 18\% in 2016 among children and adolescents aged 5-19 years \citep{WHO_Obesity_and_overweight}. Overweight or obesity in childhood is critical as it often persists into adulthood due to both physiological and behavioural factors. Indeed, childhood obesity is associated with increased risks of glucose intolerance, hypertension, dyslipidaemia, insulin resistance, and T2D in adulthood \citep{freemark2010pediatric}. Therefore, preventing childhood obesity can help disrupt the incidence of metabolic diseases in later life. Several different mechanisms such as insulin resistance, inflammation and metabolic dysregulation mediate the link between obesity and the risk of metabolic diseases. Indeed, there is a complex interplay between genetic determinants, behavioural and environmental factors which contribute to obesity. Yet, relatively little is known regarding its underlying pathophysiology. 

In this work, we investigate the complex metabolic pathways in childhood obesity, combing metabolite concentration data (as measured by NMR spectroscopy) with more traditional clinical makers measuring the growth of the children. Metabolites are small molecules that participate in metabolic reactions and are involved in biochemical pathways associated with metabolism in health and disease \citep{ellul2019metabolomics}. As the prevalence of obesity is rapidly increasing in children and adolescents, metabolomics is a powerful tool to uncover underlying biological mechanisms, to unravel genetic and environmental interactions, to identify therapeutic targets, to facilitate early detection of metabolic diseases and to monitor disease progression. Applying metabolomic techniques in relation to childhood obesity could pave a way in defining biomarkers of future metabolic risk and targets for early detection and intervention. Previous studies in adults have consistently identified metabolic signatures associated with obesity, insulin resistance and T2D. For example, previous research has reported associations between obesity and elevated plasma concentrations of amino acids such as branched-chain amino acids (BCAA, leucine, isoleucine and valine), aromatic amino acids (AA, phenylalanine and tyrosine), gluconeogenesis intermediates and glutamine metabolism which are linked to inflammation of white adipose tissue in obesity \citep{takashina2016associations, petrus2020glutamine}. Although the metabolomic literature on adults suffering from obesity, insulin resistance or T2D presents convincing results, metabolic changes related to obesity in younger populations have been poorly identified and findings are often inconsistent and different from those in adult populations \citep{balikcioglu2018metabolomic}. For instance, in contrast to  adults, a study on children and adolescents in Germany does not report association between BCAA levels and obesity \citep{wahl2012childhood}, while concentrations of medium- and long-chain acylcarnitines (C12:1 and C16:1) are reported as higher in obese as compared to normal weight children. This latter association has been replicated in adults. Moreover, further 12 metabolites (glutamine, methionine, proline, nine phospholipids) were found to be significantly altered in obese children. The identified metabolite markers are indicative of oxidative stress and of changes in sphingomyelin metabolism, in $\beta$-oxidation, and in pathways associated with energy expenditure. Contrary to adults, previous studies \citep{mihalik2012metabolomic} show that obese and diabetic children present no evidence of defects in fatty acid or amino acid metabolism as compared to their normal weight peers. In the cohort study Project Viva \citep{oken2015cohort}, BCAA concentrations have been reported to be higher in obese versus lean children aged 6--10 years \citep{perng2014metabolomic}. Similarly, it is reported \citep{butte2015global} that the concentrations of BCAAs, glutamate, lysine, tyrosine, phenylalanine, and alanine significantly increase in obese children as compared with normal weight children. However, other amino acids such as asparagine, aspartate, glycine, serine, and histidine levels decrease. These results indicate that childhood obesity influences the composition of the serum metabolome, pointing towards potential biomarkers.

The aim of this work is to  identify metabolic signatures of obesity in children with different trajectories of adiposity from 5 to 9 years of age, using
data from the Growing Up in Singapore Towards healthy Outcomes (GUSTO) prospective cohort study \citep{soh2014cohort}. Metabolome analysis is particularly  relevant in Asian populations where the risk of metabolic diseases is higher than in the western population \citep{misra2011obesity} and the GUSTO cohort study provides an optimal platform. GUSTO is a deeply phenotyped prospective cohort involving Singaporean mothers and their children, started in 2009 (pre-natal) by recruiting mothers at the first trimester of pregnancy. A wealth of information is available on both mothers and children. In this work we focus on growth profiles of children and their relationship with metabolic outcomes, as well as more traditional risk factors such as demographics and clinical biomarkers. To this end, we propose a joint model for the growth trajectories and anthropometric measures of children from birth to 9 years of age and a set of metabolites measured at age 8 years in children. The anthropometric indicators are obtained with Quantitative Magnetic Resonance (QMR) techniques \citep{chen2018body}, recording the percentages of fat and lean mass in the children's body excluding the contribution from the bones, and by height/weight measurements, used to compute the standardised body mass index (Z-BMI). These growth indicators are recorded at different time points in the children's development: every year from age 5 to 9 for the QMR measures, and at 21 unequally spaced time points for the Z-BMI. These data pose challenges to the statistical analysis given their dimension and missing rates for some of the variables, as not all subjects took part to follow-up visits. 

The main contribution of this work is to provide a joint model for three growth markers and metabolic associations, which allows for data-driven clustering of the children and highlights metabolic pathway involved in child obesity. To this end, in a Bayesian framework, we specify a joint nonparametric random effect distribution on the parameters characterising the longitudinal trajectories of obesity and the graph capturing the association between metabolites. The choice of a nonparametric random effect distribution allows for extra flexibility, heterogeneity in the population as well as data-driven  clustering of the subjects.

The paper is structured as follows: Section 2 introduces the proposed approach for the joint modelling of growth trajectories and metabolic associations; Section 3 presents posterior inference results  obtained when applying the proposed methodology to the GUSTO data, highlighting  cluster-specific growth evolution as well as differences in metabolic associations. Section 4 concludes the paper with a discussion. A Supplementary Material file is available, containing additional Figures and Tables, as well as details on the MCMC algorithm. Additionally, this file contains a description of the dataset used in the analysis.

\section{Joint modelling of growth trajectories and metabolites}\label{sec:joint-modelling-of-growth-trajectories-and-metabolites}

The main goal of the analysis is to combine information from the longitudinal responses (i.e. the growth curves) and the metabolic variables (observed only at one time point) to gain a better understanding of the children's development. In particular, we develop a joint model where the longitudinal 
outcomes are flexibly modelled via a nonparametric mixture of Gaussian Processes (GP), while we exploit tools from the Gaussian Graphical Model (GGM) literature to introduce information from metabolites and their inter-dependencies. These two components are then linked hierarchically by the specification of a suitable prior distribution.

Let $\mathcal{Y} = \{Y_t : t \in \mathbb{R}^+\}$ be a stochastic process indexed over the positive real line, in this work representing the time component, and taking values in $\mathbb{R}$. Let the realisations of such process be the vectors $\bm Y_i = (Y_{i1}, \dots, Y_{in})$, observed at times $t = t_{1}, \dots, t_{n}$ (not necessarily equidistant) for subjects $i = 1, \dots, N$. A possible modelling strategy consists of assuming a Gaussian Process to model each trajectory over time, resulting in a multivariate Gaussian likelihood distribution for the vectors $\bm Y_i$. This modelling strategy is flexible and allows for efficient computations, but would not be able to account effectively for subject heterogeneity, typical of medical studies. This is evident when observing the empirical distribution of the growth indicators, presenting skewness and heavy tails at the different observed time points, shown in Figures 1 and 2 in Supplementary Material. A possibility to overcome these limitations is to model the observations using a mixture of multivariate Gaussian distributions, adopting a flexible mixing measure which in turn involves a suitable temporal dependence structure such as the one offered by the GP. In order to do so, we extend an existing modelling strategy \citep{gelfand2005bayesian} to our context and thus assume the distribution of the vectors $\bm Y_i$ to be an infinite mixture of  multivariate Gaussians where the mixing measure is given by a Dirichlet Process (DP) prior \citep{ferguson1973bayesian}, centred around a stationary GP. The DP defines a probability measures over the space of probability distributions. A constructive definition of the DP is provided by the stick-breaking representation \citep{sethuraman1994constructive}: $P(\cdot) = \sum\limits_{j = 1}^{\infty}w_j \delta_{\psi_j}(\cdot)$,
where $\delta_{x}(\cdot)$ is the Dirac's delta measure taking value 1 at the location $x$, and 0 otherwise. The infinite sequence of locations $\{\psi_j\}_{j = 1}^{\infty}$ is an i.i.d. sample from a centering measure $P_0$, while the infinite sequence of weights $\{w_j\}_{j = 1}^{\infty}$ is constructed in the following way:
\begin{align*}\label{eq:SBweights}
    & w_j = v_j \prod_{i < j}(1 - v_i), \quad j = 2, 3, \dots, \quad w_1 = v_1, \quad v_1, v_2, \dots \iid \text{Beta}(1, \alpha) \nonumber
\end{align*}
where the mass parameter $\alpha > 0$ controls the dispersion of the process around $P_0$. An important feature of the DP making it appealing in applications is its almost sure discreteness, implying the possibility of modelling ties in the sample from this distribution, inducing a partition of the indices $\{1, \dots, N\}$. When used as a mixing measure, this implies a partition of the subjects sharing the same value of the mixing parameter. As previously pointed out \citep{gelfand2005bayesian}, this yields a flexible distribution for the vectors $\bm Y_1, \dots, \bm Y_N$, which is non-Gaussian and non-stationary, but retains the advantageous mathematical and computational properties of the GP (see Supplementary Section 5 for the details of the algorithm). This, in conjunction with the computational tractability of the DP, allows for efficient posterior inference of the proposed approach. 
In our application, we have $S = 3$ distinct processes representing the percentages of fat and lean mass in the body and the standardised body mass index (Z-BMI), and therefore we model the vectors $\bm Y^{(s)}_i = (Y^{(s)}_{i1}, \dots, Y^{(s)}_{in_s})$, for $s = 1, \dots, S$ via a GP with the following process-dependent covariance kernel:
\begin{equation}\label{eq:GP_Kernel}
    \bm K_{t_i t_j}^{s_1 s_2}\left(\sigma^2, \phi^2, \eta^2, \xi_{s_1}, \xi_{s_2}\right) = \text{Cov}\left(Y^{s_1}_{t_i}, Y^{s_2}_{t_j}\right) = \xi_{s_1}\xi_{s_2} \sigma^2 e^{-\frac{ (t_i - t_j)^2 }{\phi^2} } + \eta^2 \bm 1_{\{s_1 = s_2, t_i=t_j\}}
\end{equation}
with $t_i = 1, \dots, n_{s_1}$, $t_j = 1, \dots, n_{s_2}$, $s_1, s_2 \in \{1, \dots, S\}$ and $\eta^2$ is the nugget parameter, present only on the diagonal elements of the kernel.
The covariance kernel presents similar features to the widely-used exponential kernel, accounting for the presence of multiple-processes through the scaling factors $\xi_s$, for $s = 1, \dots, S$. In particular, it includes the positive coefficient $\sigma^2$ calibrating the amount of variability in the data, as well as $\phi^2$ regulating the impact of the distance between time points on the correlation between observations, and it is stationary since it only depends on time distance $|t_i - t_j|$. We indicate by $\bm K = \left[\bm K_{t_i t_j}^{s_1 s_2}\right]$ the covariance matrix obtained from Eq. \eqref{eq:GP_Kernel} at the observed time points. The mean function of the GP is modelled via the inclusion of the subject-specific parameters $\bm \theta_i = (\bm \theta^{(1)}_i, \dots, \bm \theta^{(S)}_i)$ of dimension $p_Y = \sum_{s = 1}^S n_s$, obtained by concatenating the random intercept vectors relative to each longitudinal process. Additionally, a regression term is included in the expression of the mean of the GP, see Eq.~\eqref{eq:Full_model}.

The second component of the data is represented by the metabolite concentrations measured at year 8. These type of data are usually modelled via multivariate distributions, most commonly Gaussian. Of particular interest in this analysis is the relationship between the observed metabolites, quantifiable by their correlation structure, with the aim of understanding the dependencies between them and their role in the activation of specific metabolic pathways. We approach this problem by modelling the vectors of metabolites borrowing from the GGMs literature. In this setting, let $\bm M = (M_1, \dots, M_{p_M}) \in \mathbb{R}^{p_M}$ be a vector of $p_M$ metabolites and let $G = (V,E)$ be a graph defined over the set of nodes $V = \{1, \dots, p_M\}$ and with edge set $E \subset \{(h,k) \in V \times V | h < k \}$ such that if there is a connection between the nodes $h$ and $k$, then $(h,k) \in E$. The graph $G$ is used to represent the correlation structure of the vector of metabolites $\bm M$, exploiting the property that two elements of the vector are conditionally independent given the rest if and only if the precision matrix is null at the corresponding position \citep{dempster1972covariance}. A zero in the precision matrix corresponds to a zero in the adjacency matrix, which results in the absence of an edge in the graph (and vice-versa, an edge in the graph corresponds to a non-zero element in the precision matrix). The vectors of metabolites are modelled using a multivariate Gaussian distribution with precision matrix $\bm \Omega_G$, whose prior distribution is defined conditionally to the graph structure $G$. The standard conditionally conjugate prior distribution is the G-Wishart \citep{Roverato2002} with $\nu$ degrees of freedom, scale matrix $\bm \Psi$ and graphical encoding $G$, denoted here as $\text{G-Wishart}(\bm \Omega_G | \nu, \bm \Psi, G)$.
The prior distribution over the graph structure is given by the product of i.i.d. Bernoulli priors on each edge with inclusion probability $d \in (0,1)$, so that $\pi(G | d) \propto d^{|E|}(1 - d)^{\binom{p_M}{2} - |E|}$, with $|E|$ being the number of edges in the graph $G$ and $\binom{p_M}{2}$ the total number of possible edges. 

As mentioned earlier, the aim of this work is to study the relationship between the growth indicators and the metabolite values jointly, still allowing for flexibility. We specify a joint prior distribution for the hyperparameters of the two sub-models, thus capturing the dependency between the longitudinal and metabolic dimensions. In particular, we specify a joint distribution on the random effect vector $\bm \theta$ in the longitudinal part of the model, on the precision matrix $\bm \Omega_G$ and embedded graph $G$. Let $\bm \psi_i = (\bm \theta_i, \bm \Omega_{G_i}, G_i)$ for $i = 1, \dots, N$ be the array of subject-specific parameters of interest. We specify a the DP prior on the arrays $\bm \psi_1, \dots, \bm \psi_N$ in order to link the two sub-models corresponding to the growth trajectories and the metabolite concentrations, obtaining:

\begin{align}\label{eq:Full_model}
	& \text{\underline{Longitudinal}:} \\
	& \bm Y^{(s)}_i | \bm \theta^{(s)}_i, \bm \beta^Y_s, \bm X^Y_i, \tau^2_s \sim \text{N}_{n_s}(\bm Y^{(s)}_i | \bm \theta^{(s)}_i + \bm \beta^Y_s \bm X^Y_i, \mathbb{I}_{n_s}/\tau^2_s) \nonumber \\
	& \bm \beta^Y = \left[\bm \beta^Y_1, \dots, \bm \beta^Y_S\right] \sim \text{MN}_{p_Y \times q_Y}(\bm \beta^Y | \bm 0, \mathbb{I}_{p_Y}, \mathbb{I}_{q_Y}) \nonumber \\
	& \tau^2_s \sim inv-gamma(\tau^2_s | 3, 2) \nonumber \\
	& \sigma^2, \phi^2, \eta^2 \sim inv-gamma(1, 1) \nonumber \\
	& \xi_1, \dots, \xi_S \sim gamma(1, 1) \nonumber \\
	&\nonumber \\
	& \text{\underline{GGM}:} \\
	& \bm M_i | \bm \beta^M, \bm X^M_i, \bm \Omega_G \sim \text{N}_{p_M}(\bm M_i | \bm \beta^M \bm X^M_i, \bm \Omega_{G_i}) \nonumber \\
	& \bm \beta^M \sim \text{MN}_{p_M \times q_M}(\bm \beta^M | \bm 0, \mathbb{I}_{p_M}, \mathbb{I}_{q_M}) \nonumber \\
	& \bm \Omega_{G_i} | \nu, \bm \Psi, {G_i}  \sim \text{G-Wishart}(\bm \Omega_{G_i} | \nu, \bm \Psi, {G_i}) \nonumber \\
	& {G_i} | d \sim \pi({G_i} | d) \nonumber \\
	&\nonumber \\
	& \text{\underline{DP}:}\\
	& \bm \psi_1, \dots, \bm \psi_N | P \iid P, \quad P \sim DP(\alpha, P_0) \nonumber \\
	& P_0(\bm \theta, \bm \Omega_G, G) = \text{GP}(\bm \theta | \bm \mu_{\bm \theta}, \bm K) \text{G-Wishart}(\bm \Omega_G | \nu, \bm \Psi, G) \pi(G | d) \nonumber
\end{align}
where $\text{N}_p(\bm Y | \bm \mu, \bm \Omega)$ is the $p$-dimensional Gaussian distribution for the vector $\bm Y$ with mean $\bm \mu$ and precision matrix $\bm \Omega$.
where the DP prior is imposed on the $p_Y$-dimensional concatenated random effects $\bm \theta_1, \dots, \bm \theta_N$ in order to introduce dependencies within and between processes via the definition of the GP kernel in Eq. \eqref{eq:GP_Kernel}, on the graph and on the precision matrix. Finally, a conjugate Gaussian prior distribution is imposed on the mean vector of the GP in the centring measure $P_0$. We specify a joint Matrix-Normal prior distribution for the column-wise concatenated matrix of coefficients $\bm \beta^Y \in \mathbb{R}^{p_Y \times q_Y}$, where $p_Y = \sum\limits_{s = 1}^S n^s$ and $q_Y$ is the number of covariates used. The prior distribution for the matrix of coefficients $\bm \beta^Y$ has zero mean matrix and identity covariance matrices. Notice that this setting implies a different regression coefficient being estimated at each observed time point for each covariate included in the model, allowing to capture changes in the temporal effect of risk factors. We write $gamma(x | a, b)$ and $inv-gamma(x | a, b)$ to denote the Gamma and the inverse-Gamma distribution for $x > 0$ with means $a/b$ and $b/(a - 1)$, respectively. We allow for covariate effects on the metabolite concentrations via the terms $\bm \beta^M \bm X^M_i$ and use a prior specification for the matrix of coefficients analogous to the one for $\bm \beta^Y$.

The main contribution  of the proposed approach is the ability to cluster individuals based on their growth profiles and metabolic associations through joint modelling of longitudinal and multivariate markers. Modelling of multiple graphs has been proposed before in the Bayesian framework  \citep{peterson2015bayesian, tan2017bayesian, shaddox2020bayesian} with groups specified a priori, but not in the context of graph-based (unsupervised) clustering, which is achieved by our modelling strategy. To the best our knowledge, this approach has not been proposed in the statistical literature before. This modelling choice implies that the base measure $P_0$ of the DP is a mixed measure, due to the presence of the graph structure in the sample from the DP. In particular, $P_0$ is defined on the product space $\mathbb{R}^{p_Y} \times \mathbb{P}_G \times \mathcal{G}_{p_M}$, where $\mathcal{G}_{p_M}$ represents the space of all possible graphs of dimension $p_M$. In general, when using the DP, the base measure $P_0$ is chosen to be non-atomic, allowing for the computation of the predictive distributions. However, examples of applications requiring a mixed base measure in the specification of the DP are found in the literature \citep{dunson2008bayesian, guindani2009bayesian}. As reported in existing work \citep{canale2017pitman}, the mixed measure setting is not problematic in the case of the DP, since the predictive distributions remain unchanged, and thus posterior inference via a P\'{o}lya urn algorithm can still be achieved, and we exploit this result.
One of the main features of this approach is that the unique values associated with the clusters are not necessarily different, due to the mixed nature of the centering measure $P_0$. For instance, in our setting, ties between the graph structures associated with the clusters can be observed (but not among the random effects or precision matrices).

It is common (especially in medical/epidemiological research) that the set of metabolites $\bm M$ is used as predictor, often in a regression setting. From this perspective, the propose model shows similarities with existing Bayesian nonparametric literature on product partition models with covariates (PPMx) \citep{muller2011product}. In particular, it can be shown that the marginal distribution of the random partition induced by the DP measure can be factorised in terms including the covariates (i.e., the metabolites) within each cluster. Additional details are reported in Supplementary Section 1.
In principle, this would allow us to devise an algorithm similar to the one proposed in the original PPMx models. In practice, due to its computational burden, such approach is unfeasible for graphs of even moderate sizes. As such, we opt for a conditional algorithm (see Supplementary Section 5), which does not marginalise over the random measure. The algorithm is based on Metropolis-within-Gibbs sampling, with adaptive steps for those parameters which are non-conjugate in the proposed model. The update of the DP parameters follows a P\'{o}lya Urn scheme for non-conjugate models, adjusted for the presence of the non-conjugate graphical structure. The updates of the graph and precision matrix within each cluster are tackled using the Birth-and-Death algorithm of \citep{mohammadi2015bayesian}.

\section{Posterior inference}
In this Section we present the application of the proposed modelling strategy to the data from the GUSTO cohort. The longitudinal data are composed of $N = 227$ fat and lean percentages measured at years 5 to 9 of the children, together with the Z-BMI values from birth to year 9 at non-equally spaced time points, such that $n_1 = n_2 = 5$ and $n_3 = 21$. The size of the concatenated vectors of growth indicators is $p_Y = 31$. The list of $p_M = 35$ metabolites measured at year 8 for the same subjects is reported in Supplementary Table 1. We apply a logit transformation to the fat and lean percentages, in order to map the observations to the real line, while the Z-BMI is standardised. Many of the metabolites present different ranges of values, skewness, and non-normality. In order to correct for these features, we apply a Box-Cox transformation to each metabolite individually and then standardise the observations component-wise.

Additionally to longitudinal and metabolic information, demographic variables and other clinical characteristics, such as ethnicity, pre-pregnancy maternal BMI, findings from oral glucose tolerance test (OGTT) and gender of the child, are available. The full list of covariates  is reported in Supplementary Table 2. The same set of covariates is used in both regression components, therefore $\bm X^Y = \bm X^M$ and $q_Y = q_M = 14$, without any intercept term. The covariates present a relatively low percentage of missing values, as reported in Supplementary Table 2, which are imputed using the R package \texttt{mice}. After imputation, the continuous covariates are standardised.

We fix the hyperparameters for $\tau^2_s$, $\sigma^2$, $\phi^2$, $\eta^2$ and $\xi_s$ for $s = 1, \dots, S$ so that their prior means and variances are both equal to 1; we set $d = 2 / (p_M - 1) \approx 0.06$ inducing sparsity in  the graph structure \citep{jones2005experiments}; we fix the mass parameter of the DP \citep{jara2007dirichlet} to $\alpha = 0.18$, yielding $\mathbb{E}(K_N) \approx 2$ and $Var(K_N) \approx 1$; the hyperparaemters of the centering measure $P_0$ are set as $\bm \mu_{\bm \theta} = \bm 0$, $\nu = p_M + 2 = 37$ and $\bm \Psi = 10 \mathbb{I}_{p_M}$. The MCMC algorithm is run for 50000 iteration after an initial burn-in period of 100 iterations used to initialise the adaptive steps. Then, after a burn-in of 40000 iterations, 5000 iterations are saved with a thinning of 2.

\subsection{Posterior Inference on Clustering Allocation}

\begin{figure}[ht]
	\centering
	\includegraphics[width=1\textwidth]{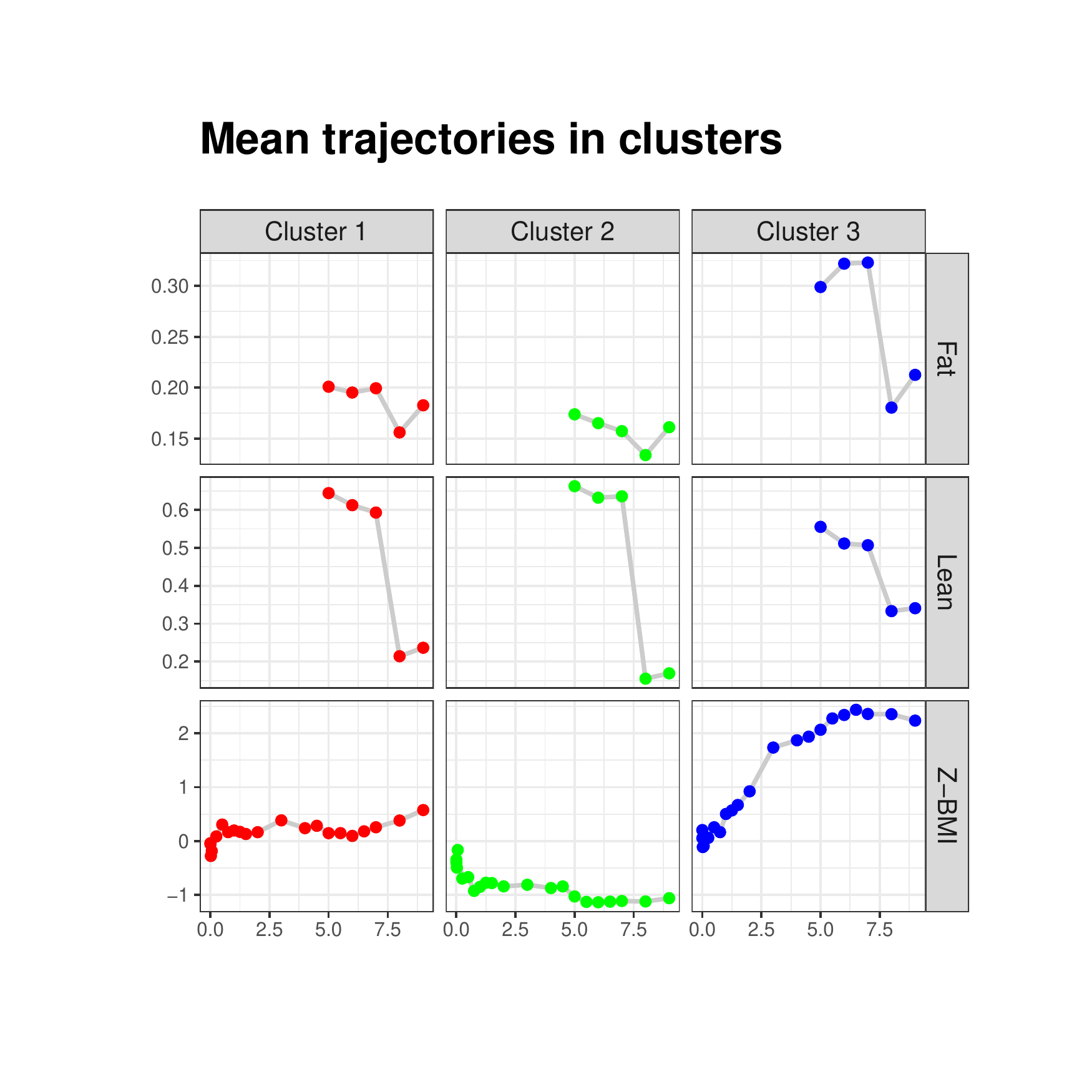}
	\caption{Posterior mean trajectories within each cluster identified by the Binder's partition. Each row represent a growth indicator (Fat/Lean/Z-BMI), while each column refers to a cluster.}
	\label{fig:Trajectories_inClust_mean}
\end{figure}

An advantage of the proposed approach is the ability of provide posterior inference on the clustering of the subjects, therefore allowing for the identification of groups of children characterised by specific growth trajectories and metabolic associations. As posterior estimate of the random partition $\rho_N$, we report the clustering configuration minimising the Binder's loss function \citep{binder1978bayesian}, which corresponds the expected loss derived from the two possible misclassification errors, occurring when a pair of subjects is erroneously clustered together or separately. The use of Binder's criterion yields a partition composed of three clusters of sizes 124, 71 and 32, respectively. We refer to it as the Binder partition, and label the clusters by their decreasing size. The number of clusters identified by Binder's method coincides with the posterior mode of the random variable counting the number of clusters, reported in Figure 3 of Supplementary Material, together with the posterior co-clustering probability for each pair of subjects. The results show little uncertainty in the distribution of the number of clusters and cluster assignment. In order to visualize the data and their estimated partition, we display the mean of the longitudinal growth data within each cluster in Figure \ref{fig:Trajectories_inClust_mean}. As it is evident from the top panel of the Figure, the longitudinal growth patterns display an intuitive clustering structure, separating children with low fat and high lean percentages from those with high fat and low lean percentages. The biggest cluster is composed of children with moderate values of fat and lean percentages. The Z-BMI curves follow a similar pattern, but only after 15 months of age.

\begin{figure}[ht]
	\centering
	\includegraphics[width=0.75\textwidth]{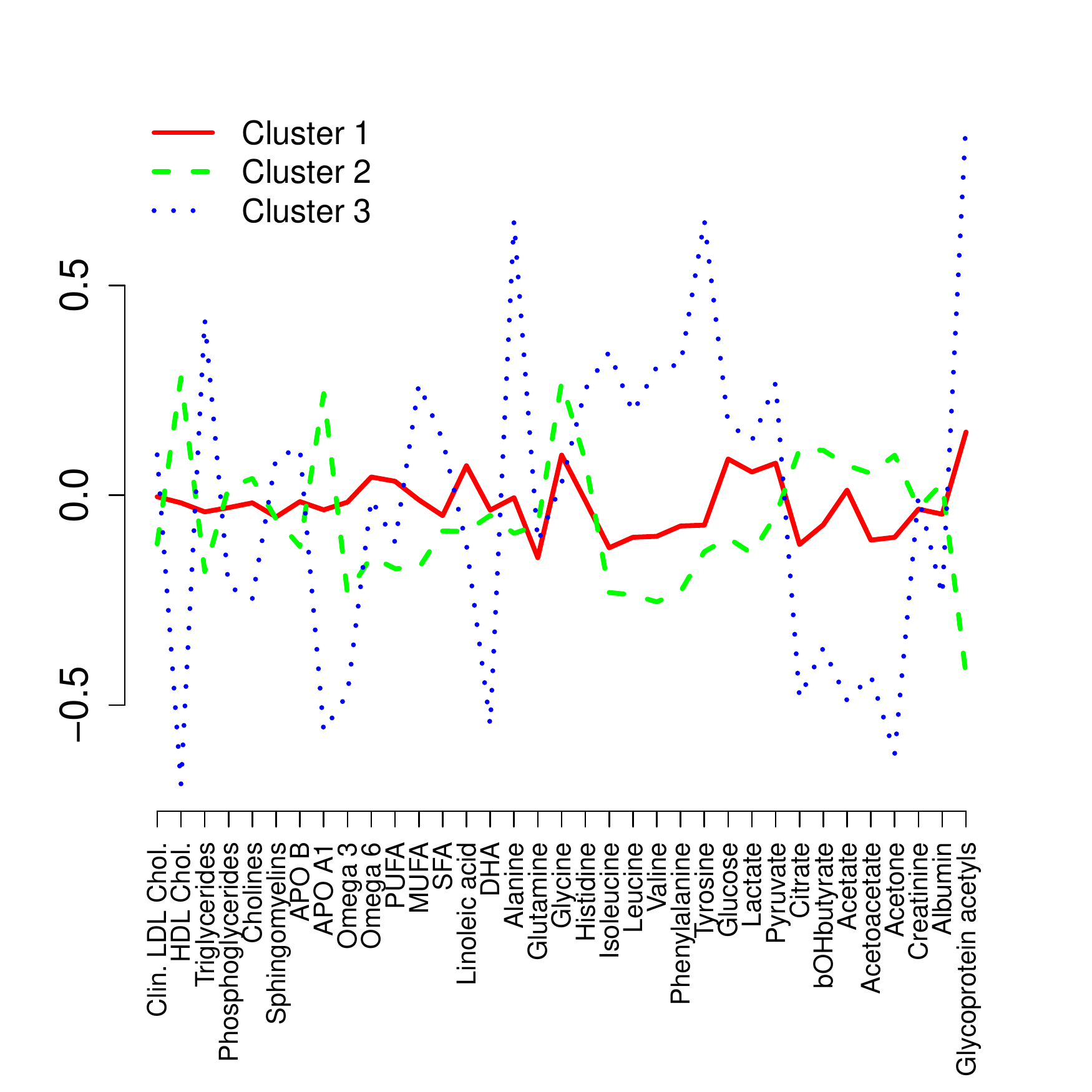}
	\caption{Empirical  mean of metabolite concentration within each cluster identified by the Binder's partition.}
	\label{fig:Metabolites_inClust_mean}
\end{figure}

As exploratory analysis, we also look at the empirical mean of the metabolite concentrations within each cluster, shown in Figure \ref{fig:Metabolites_inClust_mean}. The three clusters exhibit different mean patterns, highlighting differences in the three groups also on a metabolic level. It is evident that the mean level of almost all metabolites in  Cluster 1 is around zero, while in the smaller clusters the distributions are centred away from zero and often in opposite directions between them, supporting the hypothesis that they capture different metabolic mechanisms.

Summarizing the posterior distribution of the latent variables $\bm \psi_1, \dots, \bm \psi_N$ is not trivial, due to the existence of label-switching problems arising when working in a Bayesian nonparametric setting. Therefore, in order to understand the results of the clustering analysis, we run an additional MCMC chain, with the same number of iterations, after fixing the random partition $\rho_N$ to be equal to the Binder partition. By doing so, we are able to provide the posterior distribution of the values of $\bm \psi^\star$ within each of the three clusters. We show in Figure \ref{fig:G_all} the posterior estimates of the graph structures within each of the three clusters (the corresponding estimates of the precision matrices are shown in Supplementary Figure 4). The estimates are the median graphs, obtained selecting the edges whose posterior inclusion probability is greater than 0.5 \citep{barbieri2004optimal}. We observe that the number of estimated connections in the graphs is highest in Cluster 1, as well as the intensity of the entries of the corresponding precision matrix (see Supplementary Figure 4). In all clusters, we can identify a group of metabolites linked together, corresponding to fatty acids, phosphogrlycerides, apolipoproteins and cholesterol (see Figure \ref{fig:G_all} and Supplementary Figure 4, top left corners), while associations between smaller groups of metabolites involving some amino acids and ketone bodies show different patterns in the three clusters. We provide a discussion and suggest a biological interpretation of such differences in Section \ref{sec:differential-network-analysis}.

\subsection{Posterior Inference on Regression Coefficients}

We now discuss posterior inference on the regression coefficients for the three responses. As described in Section \ref{sec:joint-modelling-of-growth-trajectories-and-metabolites}, we estimate the effect of the covariates on each growth process at different time points. 
We report in Supplementary Figures 5, 6 and 7 the posterior means and 95\% credible intervals (CI) for the entries of the matrix $\bm \beta^Y$. We consider as relevant those predictors whose 95\% CI does not contain the value zero, highlighted in red in the Figures. Interestingly, the covariates which influence the fat and lean percentages at most of the five time points include gestational age, gender of the child, maternal pre-pregnancy BMI and ethnicity, confirming existing results obtained from the same cohort \citep{ong2021cardiometabolic}. Moreover, some covariates have different effects across time, such gender and highest education degree of the mother. Similar results on the effect of these covariates on the evolution of the Z-BMI trajectories are also reported in Supplementary Figure 7. Posterior estimates of $\bm \beta^M$ are shown in Supplementary Figure 8.

\subsection{Differential network analysis}\label{sec:differential-network-analysis}

\begin{figure}[ht]
	\centering
	\subfloat[]{\includegraphics[width=0.35\textwidth]{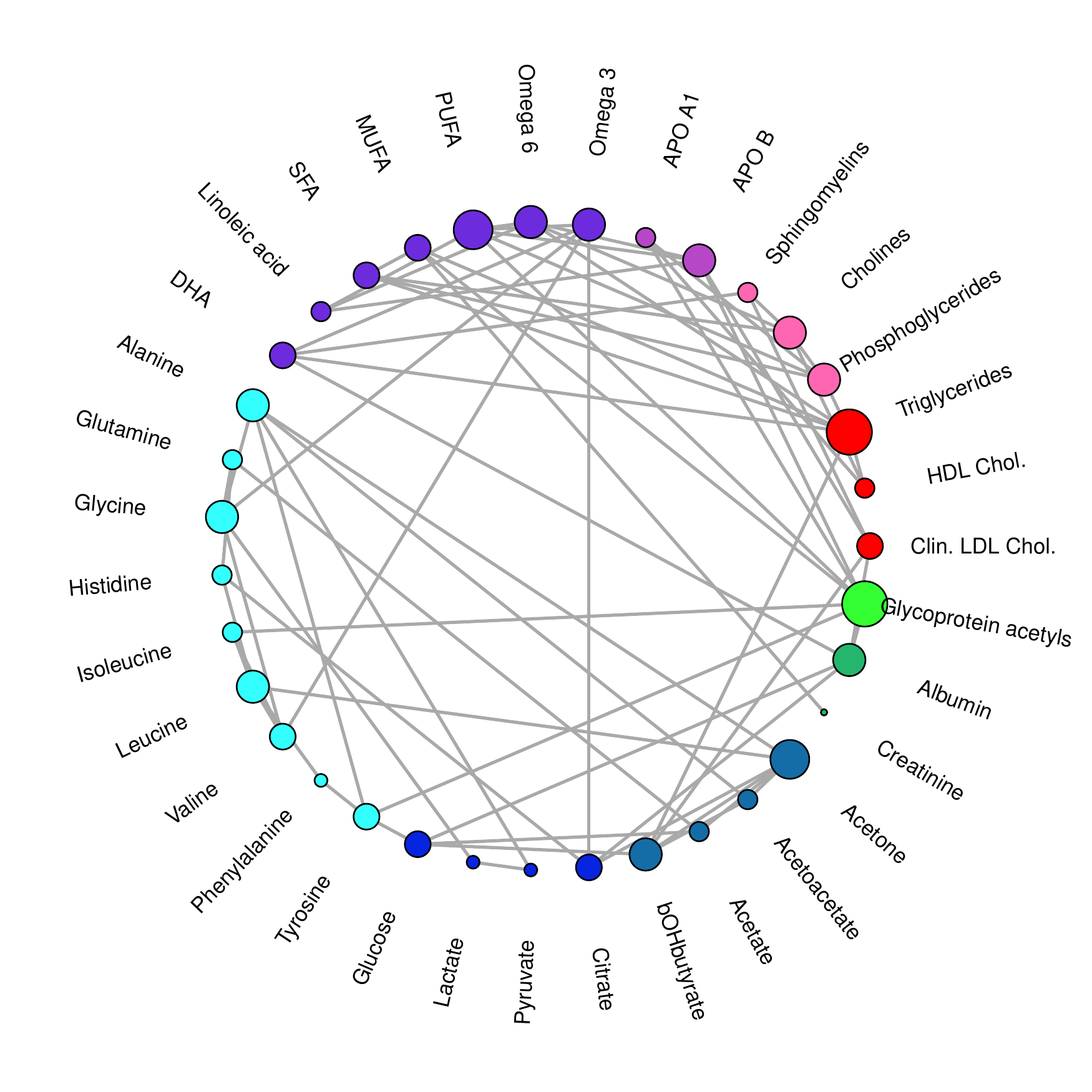}}
	\subfloat[]{\includegraphics[width=0.35\textwidth]{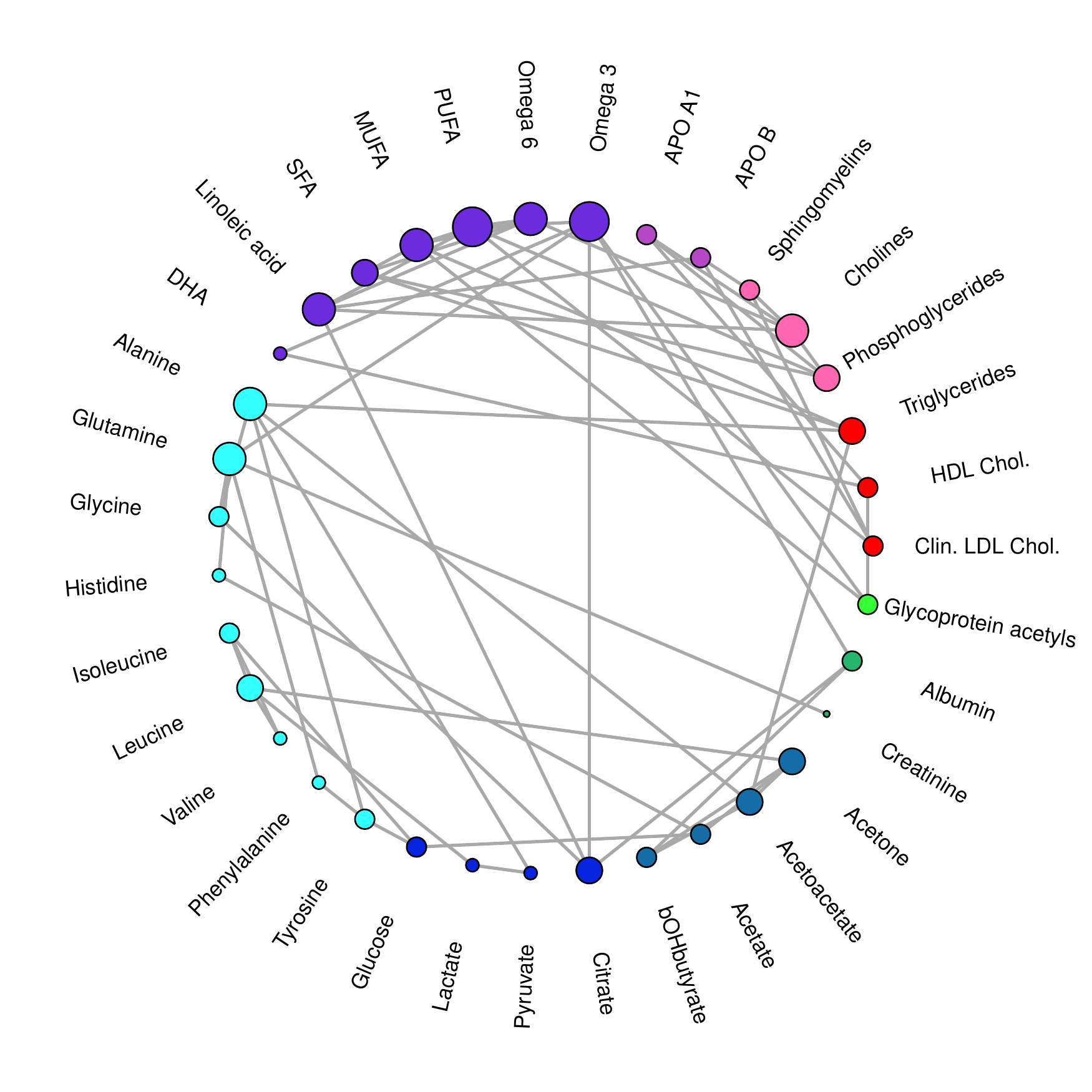}}
	\subfloat[]{\includegraphics[width=0.35\textwidth]{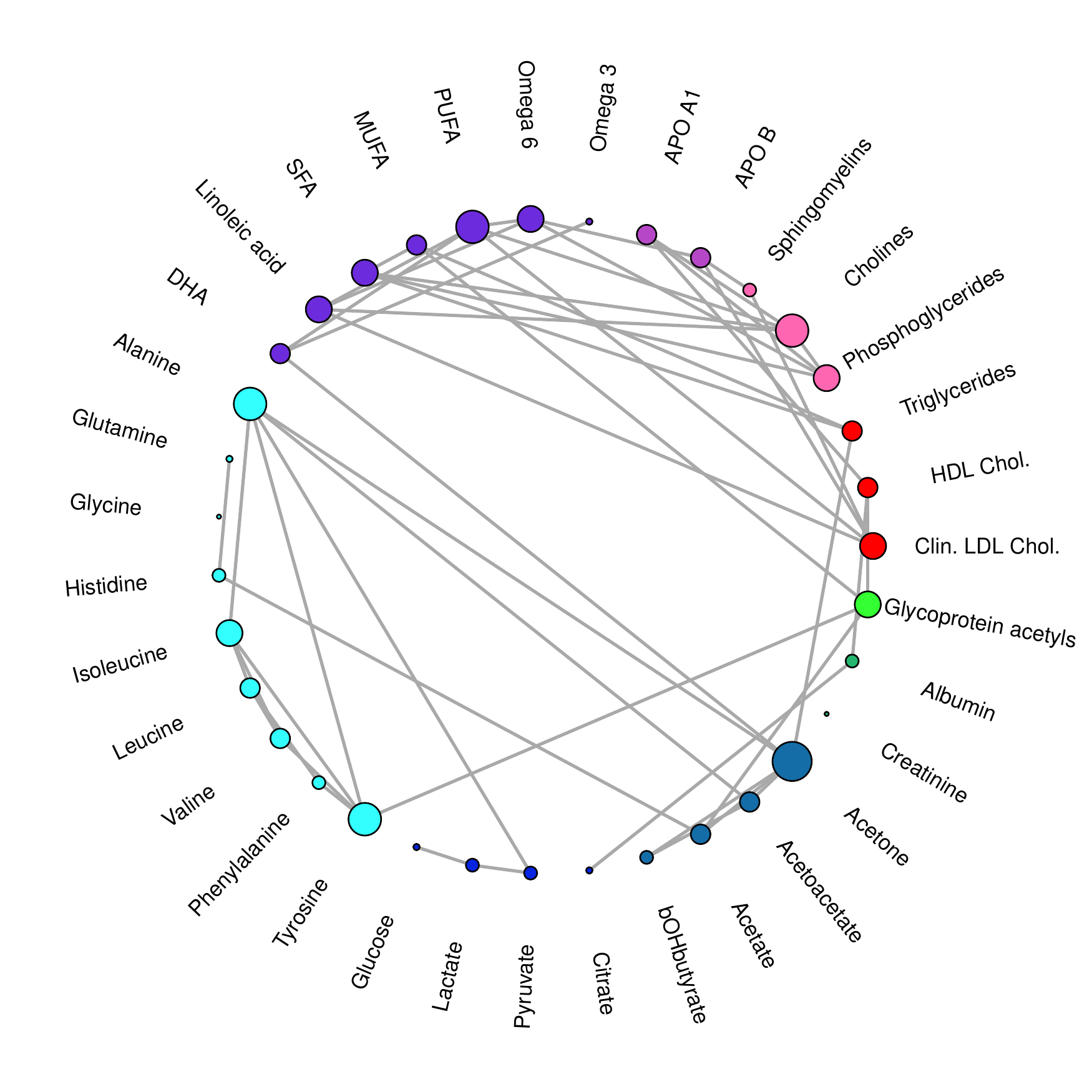}}\\
	\centering
	\includegraphics[width=0.75\textwidth]{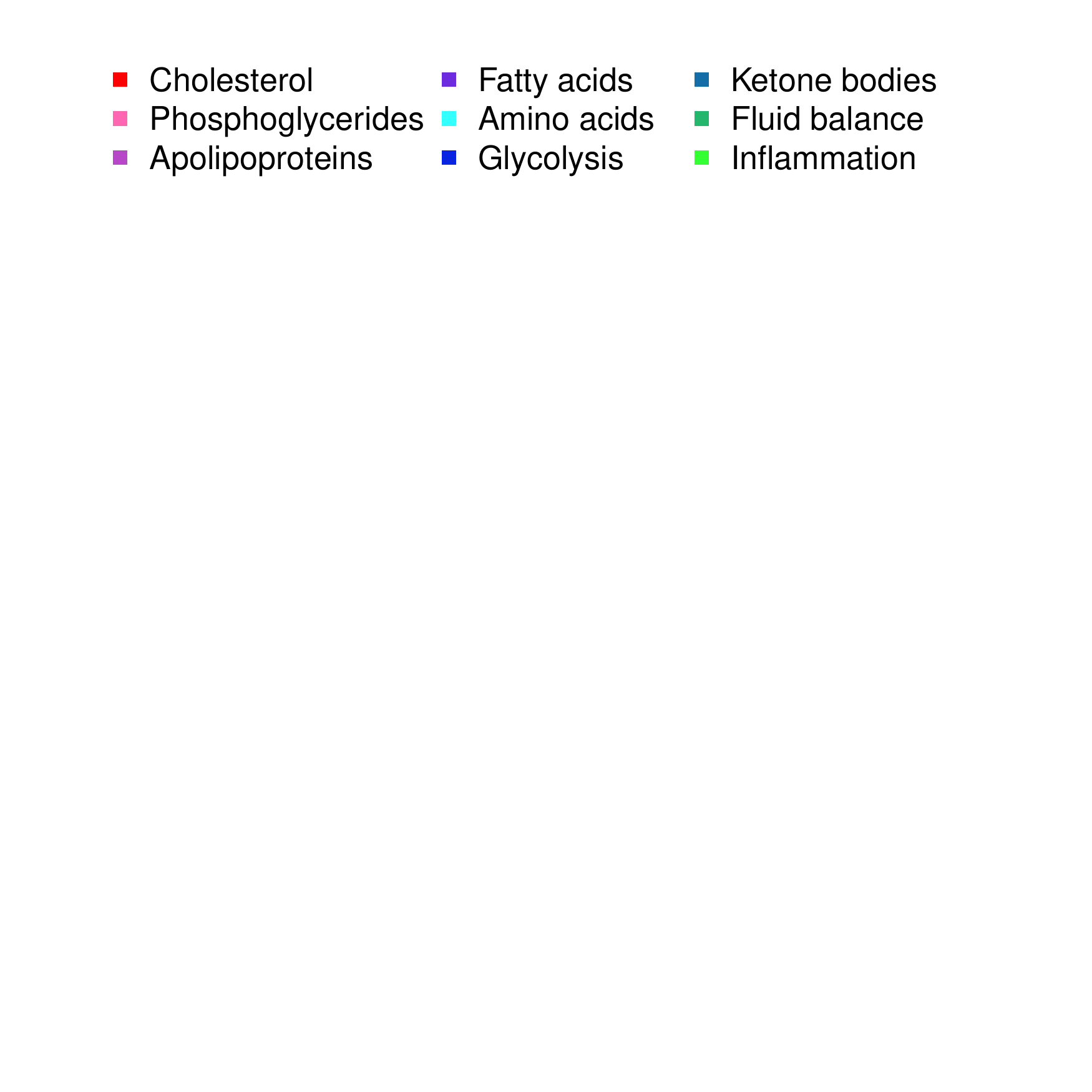}	
	\caption{Posterior median graph within each cluster. The colours indicate different chemical classes by \cite{bell2020early}.}
	\label{fig:G_all}
\end{figure}

\begin{figure}[ht]
	\subfloat[Differential network]{\includegraphics[width=0.5\textwidth]{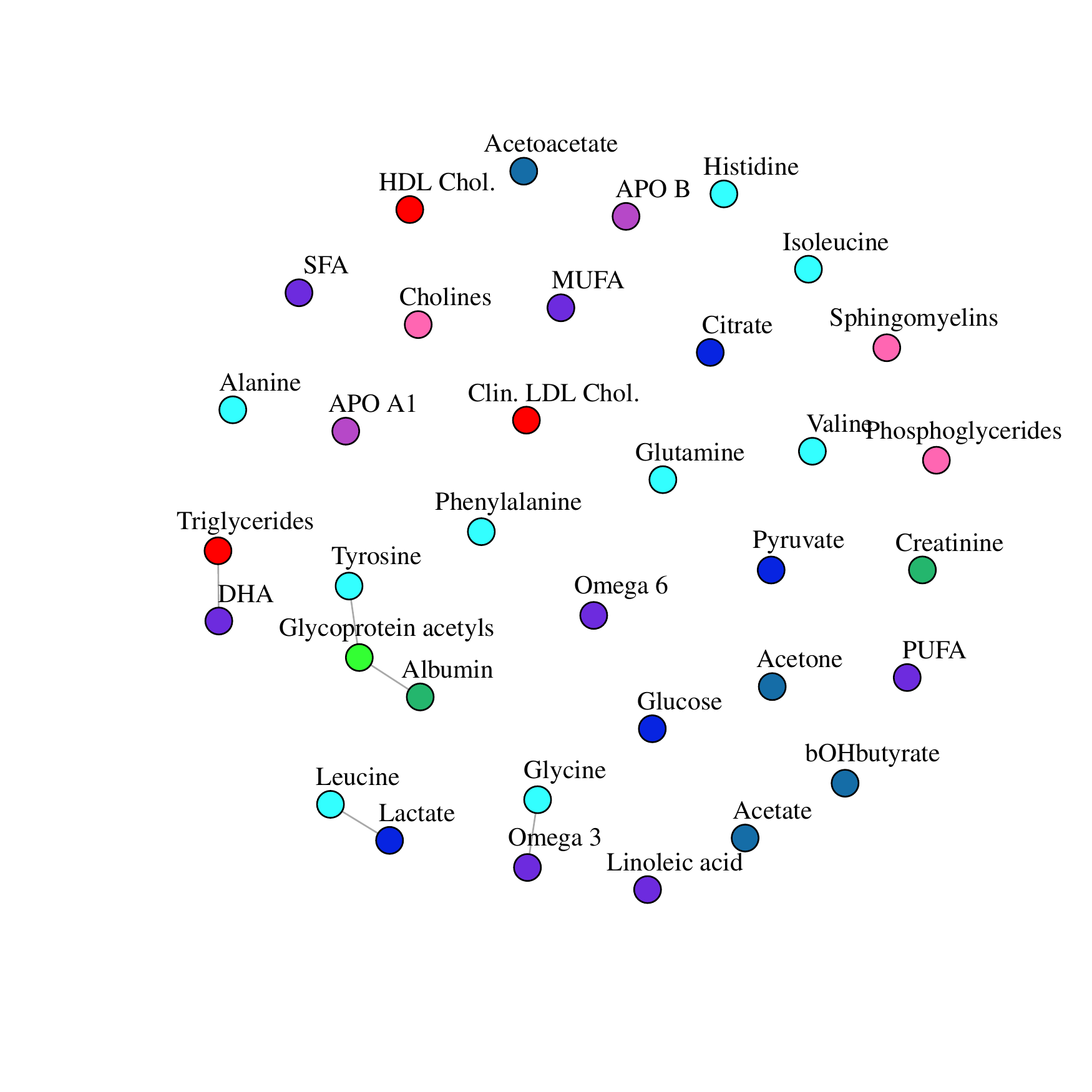}}
	\subfloat[IPA]{\includegraphics[width=0.55\textwidth]{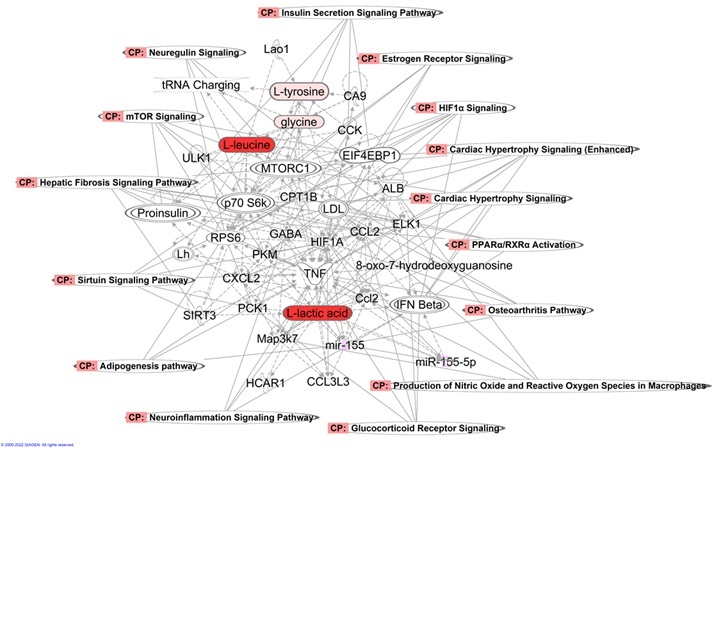}}
	\caption{Posterior differential network and result of IPA between Clusters 1 and 2. The threshold for edge inclusion is set to 0.9. The colours in panel (a) indicate different chemical classes by \cite{bell2020early}.}
	\label{fig:DiffNet_12}
\end{figure}

\begin{figure}[ht]
	\subfloat[Differential network]{\includegraphics[width=0.5\textwidth]{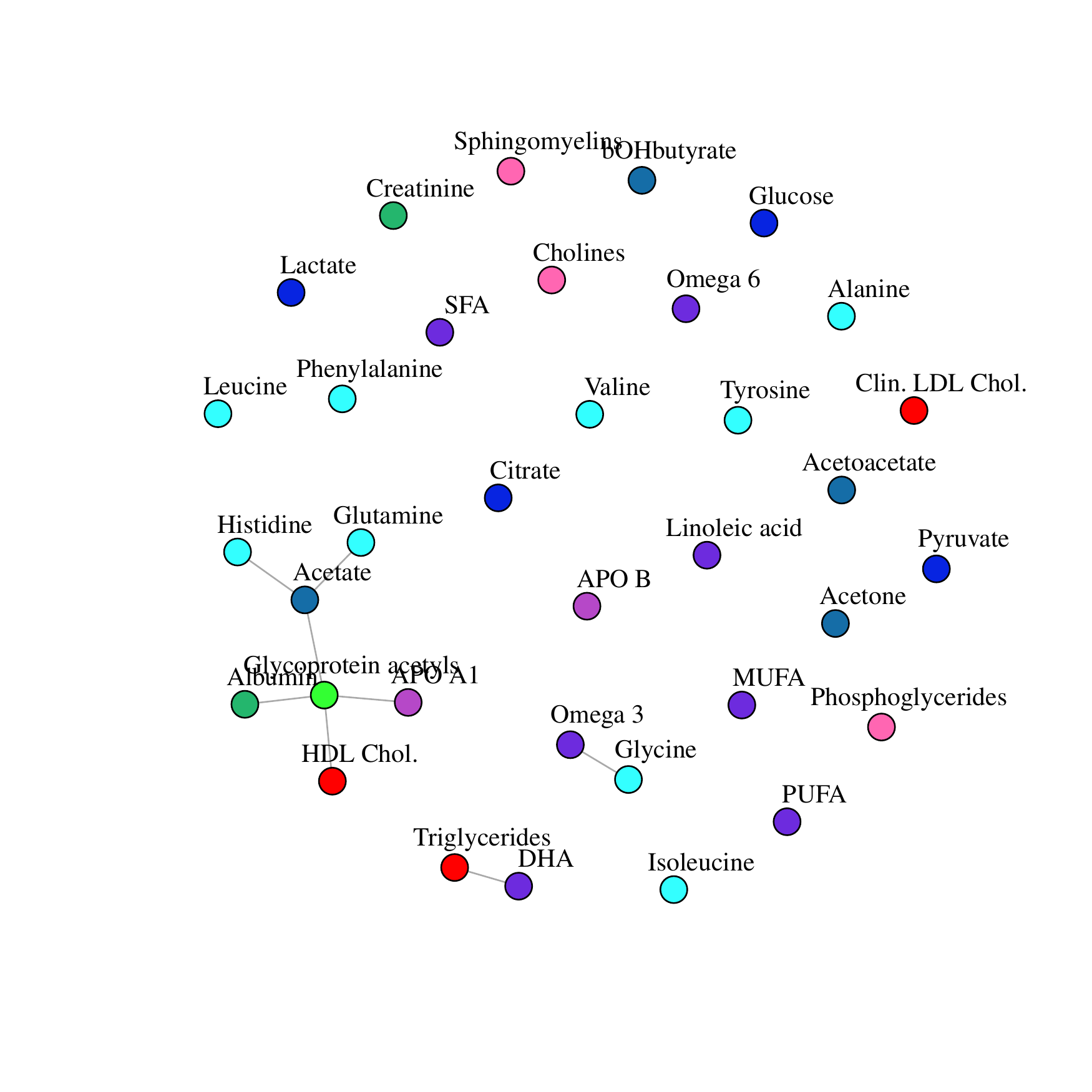}}
	\subfloat[IPA]{\includegraphics[width=0.55\textwidth]{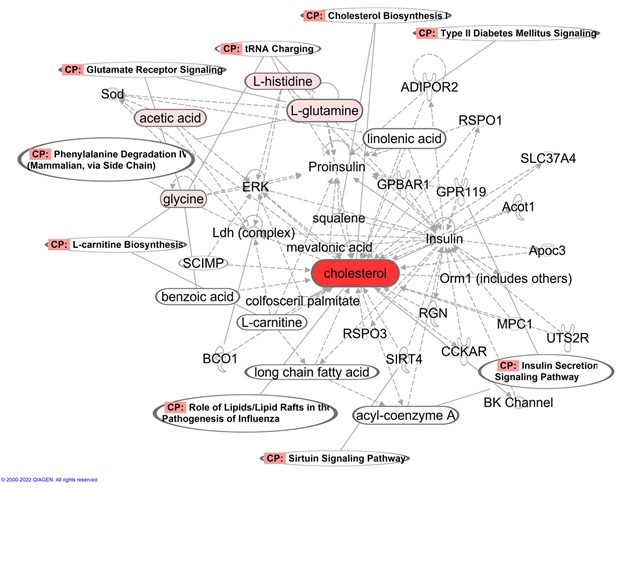}}
	\caption{Posterior differential network and result of IPA between Clusters 1 and 3. The threshold for edge inclusion is set to 0.9. The colours in panel (a) indicate different chemical classes by \cite{bell2020early}.}
	\label{fig:DiffNet_13}
\end{figure}

\begin{figure}[ht]
	\subfloat[Differential network]{\includegraphics[width=0.5\textwidth]{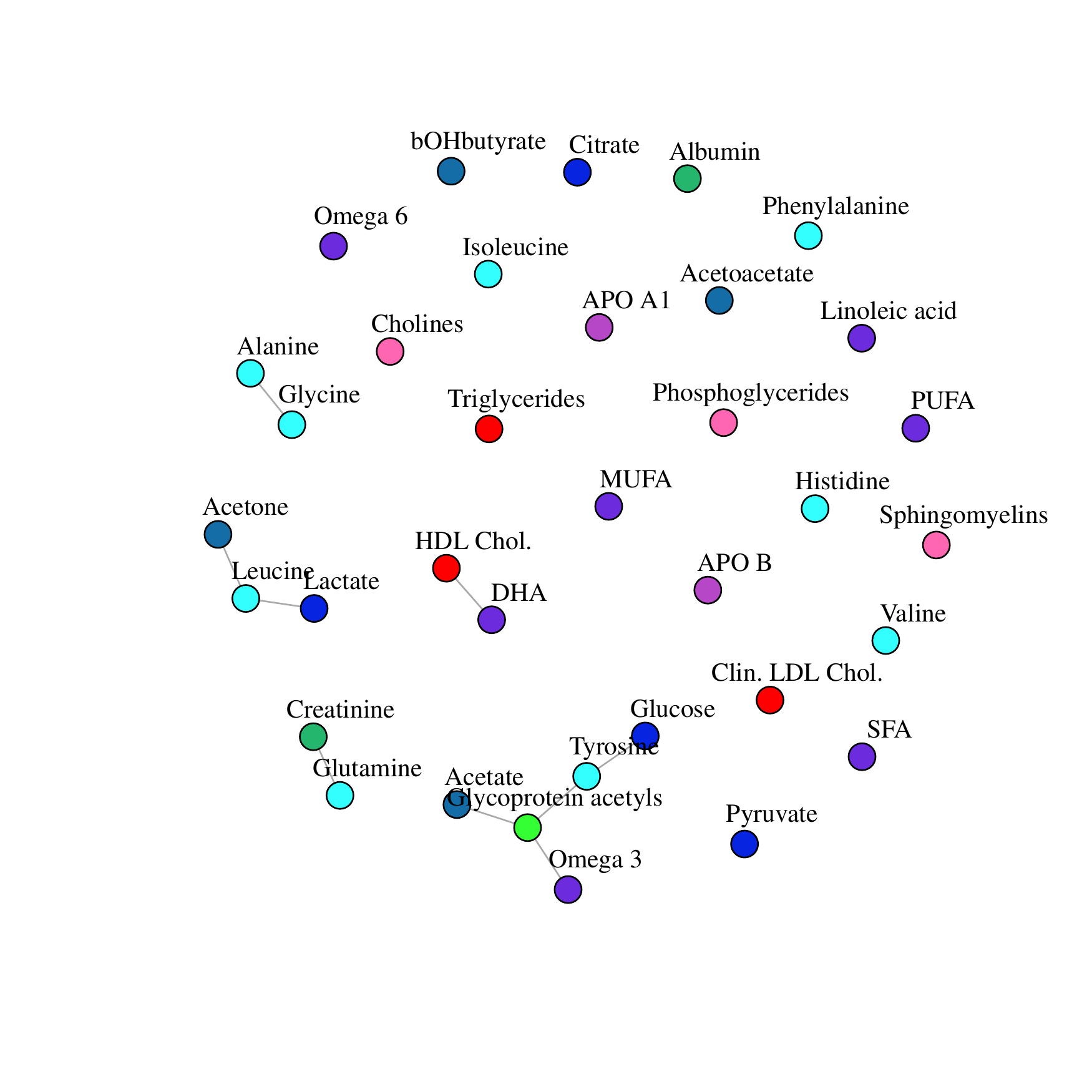}}
	\subfloat[IPA]{\includegraphics[width=0.55\textwidth]{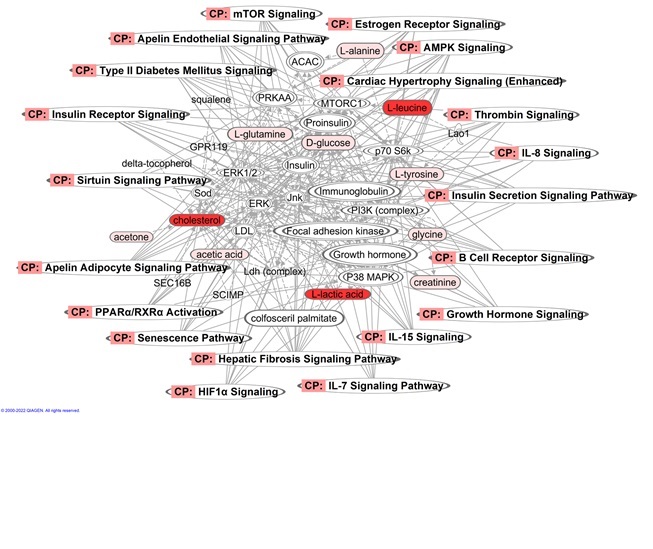}}
	\caption{Posterior differential network and result of IPA between Clusters 2 and 3. The threshold for edge inclusion is set to 0.9. The colours in panel (a) indicate different chemical classes by \cite{bell2020early}.}
	\label{fig:DiffNet_23}
\end{figure}

To quantify the differences between the cluster-specific networks presented in Figure \ref{fig:G_all}, we estimate a \textit{differential network} \citep{valcarcel2011differential, tan2017bayesian}, providing an approach based on the joint posterior distribution of the graph structures within each cluster to establish whether the differences among the cluster-specific networks are statistically relevant. We perform three pair-wise comparisons of the networks characterising the three clusters estimated by minimising the Binder's loss function. A differential network only shows those connections for which the absolute difference between the posterior edge inclusion probabilities of two graphs is greater than 0.9. Specifically, for two clusters $k_1$ and $k_2$ we require that $|\hat{\pi}^{k_1}_{ij} - \hat{\pi}^{k_2}_{ij}| > 0.9$, where $\hat{\pi}^{k}_{ij}$ is the posterior inclusion probability of the edge between nodes $i$ and $j$ in cluster $k$, estimated using the MCMC output obtained after fixing the random partition $\rho_N$ to the Binder partition, as previously done in the context of cluster-specific network estimation (see Figure \ref{fig:G_all}). The resulting differential networks are shown in Figures \ref{fig:DiffNet_12}, \ref{fig:DiffNet_13} and \ref{fig:DiffNet_23} (left panels). The differential networks are characterised by a different number of edges. However, there are key metabolites common to all three differential networks: (i) some amino acids such as glycine; (ii) glycoprotein acetyls; (iii) docosahexaenoic acid (DHA, an omega 3 fatty acid); (iv) lipids (HDL and triglycerides). Albumin is involved only in the first two differential networks, while acetate only in the last two.

\begin{figure}[ht]
	\centering
	\includegraphics[width=1\textwidth]{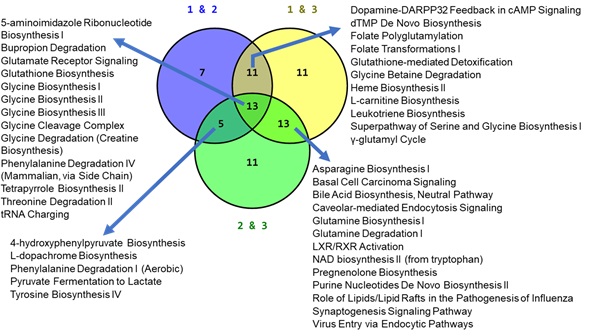}
	\caption{Venn diagram of the results of the IPA analysis performed on each differential network resulting from the pair-wise comparisons between the graphical structures estimated within each cluster. Each circle in the diagram refers to one comparison. The numbers in the diagram indicate the number of unique pathways found to be statistically significant via IPA.}
	\label{fig:IPA_Venn_1}
\end{figure}
\begin{figure}[ht]
	\centering
	\includegraphics[width=1\textwidth]{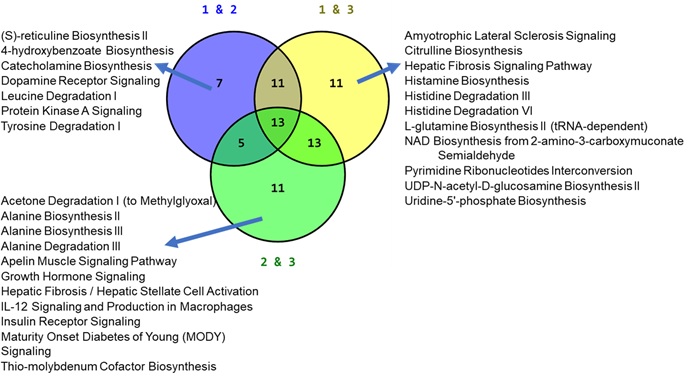}
	\caption{Venn diagram of the results of the IPA analysis performed on each differential network resulting from the pair-wise comparisons between the graphical structures estimated within each cluster. Each circle in the diagram refers to one comparison. The numbers in the diagram indicate the number of unique pathways found to be statistically significant via IPA.}
	\label{fig:IPA_Venn_2}
\end{figure}

The roles of the metabolites involved in the differential networks can be further explored by performing an Ingenuity Pathway Analysis (IPA) \citep{kramer2014causal}. IPA generates network maps between molecules (see Figures \ref{fig:DiffNet_12}(b), \ref{fig:DiffNet_13}(b) and \ref{fig:DiffNet_23}(b)), and compares them with multiple metabolic pathways (i.e. the linked series of chemical reactions occurring within a cell) available in the literature, assigning a score to each comparison. The scores are generated based on the negative logarithm of the significance level  obtained by performing Fisher's exact hypergeometric test when comparing the estimated differential network and the known pathways in the IPA library. For canonical pathway analysis, values of $-\log(\text{p-value}) > 2$ are used to detect significant activation. Figures \ref{fig:IPA_Venn_1} and \ref{fig:IPA_Venn_2} show a summary of the pathways identified as significant via the IPA methodology. We identify a total of 13 pathways through IPA that are common between the three differential networks, denoted by 1\&2, 2\&3, and 1\&3 in the Figures. Of the 13 pathways, key ones are tRNA charging, biosynthetic pathways for glycine, glutamate receptor signalling as well as degradation of the aromatic amino acid phenylalanine. The commonality among the three groups indicates an active amino acid biosynthesis machinery with the initiation from tRNA charging which is a requisite for translation and transcription of protein biosynthesis through the binding of amino acids. It has been previously shown that altered tRNA aminoacylation, modification and fragmentation are associated with $\beta$-cell failure, obesity and insulin resistance \citep{arroyo2021trna}. All the amino acid pathways common to the three differential networks are associated with obesity as well as metabolic syndrome. In Figure \ref{fig:IPA_Venn_2}, comparison between Clusters 1 and 2 (blue) highlights seven unique pathway via the IPA \ref{fig:IPA_Venn_2}, referring to the comparison between normal and low Z-BMI trajectories (see Figure \ref{fig:Trajectories_inClust_mean}). Prominent among them are leucine and tyrosine degradation pathways, involve in catecholamine biosynthesis. The latter has been previously shown to be related to obesity in children, where catecholamine resistance might promote insulin signalling in adipose tissue thus leading to the increase in lipogenesis \citep{qi2016obesity}. Another key pathway is the dopamine receptor signalling, which could be associated with the behavioural pattern towards food intake \citep{benton2016meta}. Dopamine receptor were also reported as the neurotransmitter biomarker in research on obesity \citep{dang2016associations}. The comparisons between differential networks 1\&3 (green) and 2\&3 (yellow) show clear patterns of insulin metabolism as well as activation of signalling pathways related to dysglycemia. The pattern could provide evidence of early events leading to insulin resistance as well as transition to a hyperglycemic state and onset of obesity as evident from IL-12 \cite[Interleukin 12,][]{nam2013impact}, apelin (a peptide) \citep{dray2008apelin} and growth hormone (GH) signalling \citep{hogild2019growth}. Studies show IL-12 family cytokines as prospective regulators that could cause insulin resistance due to obesity in tissues and plasma \citep{nam2013impact}. Furthermore, comparison between Clusters 2 and 3 showed primarily amino acid biosynthesis, alanine biosynthesis and degradation along with insulin receptor signalling and maturity onset of diabetes, while 1\&3 had eleven pathways, mostly degradative in nature as well as hepatic fibrosis signalling pathway. Comparing common pathways between the differential networks 1\&2 and 2\&3 shows only five common pathways: 4-hydroxyphenylpyruvate biosynthesis, L-dopachrome biosynthesis, phenylalanine degradation I (Aerobic), pyruvate fermentation to lactate and tyrosine Biosynthesis IV. Phenylalanine degradation I (Aerobic), indicates biosynthesis of tyrosine, a feeder molecule for acetoacetate involved in the synthesis of Acetyl (acetyl coenzyme A), which is important for dietary intake and energy balance.  The intersection between the set of pathways highlighted by comparisons 1\&2 and 1\&3 shows folate metabolism and L-carnitine biosynthesis (Figure \ref{fig:IPA_Venn_1}). The metabolic signalling is characteristic of a transition from normal BMI to an obese phenotype. This might indicate a transition from increase in amino acid biosynthesis/degradation and following appearance of lipid biosynthesis. A list of the metabolites identified as commonly differentially expressed between the three Clusters and associated pathways of activation is reported in Table \ref{tab:Pathways}.

\begin{table}
	\caption{Metabolites identified as commonly differentially expressed between the three clusters and associated pathways of activation derived from existing literature.}
	\label{tab:Pathways}
	\begin{center}	
		\begin{tabular}{ll}
			Metabolite & Primary Pathway \\\hline
			Tyrosine & Aromatic amino acid metabolism \\
			Leucine	& BCAA metabolism \cite{roberts2020blood} \\
			Alanine & Gluconeogenesis \cite{roberts2020blood} \\
			Glycine & Glutathione metabolism \cite{roberts2020blood} \\
			Glycoprotein acetyls & Chronic inflammation \cite{ritchie2015biomarker} \\
			DHA & Energy expenditure \\
			& Lipid catabolism \cite{kuda2017bioactive} \\
			& Anti-inflammatory pathways \cite{poudyal2011omega} \\
			Acetate & Ketogenesis, TCA cycle \cite{fletcher2019impaired} \\
			& Energy expenditure fat utilization \cite{canfora2017acetate} \\
			HDL, Triglycerides & Fatty acid metabolism
		\end{tabular}
	\end{center}
\end{table}

\section{Conclusions}
We propose a Bayesian semiparametric model enabling clustering of subjects based on both longitudinal trajectories and patterns of metabolic association. The work is motivated by a study on early mechanisms of obesity, but has wider applicability. Excess bodyweight is one of the leading risk factors contributing to the overall disease burden worldwide \citep{mcmillen2009early}. Childhood obesity is one of the major health problems in western countries and it is increasingly affecting Asian countries. The excessive accumulation of adipose tissue causes inflammation, oxidative stress, apoptosis and mitochondrial dysfunctions, leading to the development of severe co-morbidities including type-2 diabetes mellitus, liver steatosis, cardiovascular and neurodegenerative diseases which can develop early in life \citep{faienza2019mechanisms}.

Our analysis has identified potential key pathways in obese children in order to explore possible molecular mechanisms associated with child obesity (see Table \ref{tab:Pathways}). We identified 13 metabolic pathways common to the three differential networks, the majority of which involves amino acids. An analysis of these associations reveals multiple biochemical pathways such as aromatic amino acid metabolism, branched-chain amino acid metabolism, glutathione metabolism, gluconeogenesis, tricarboxylic acid cycle, anti-inflammatory pathways and lipid metabolism. The analysis shows comprehensive initiation of amino biosynthesis as well as precursor molecule degradation, NAD biosynthesis, TCA cycle responsible for providing feeder molecules to sustain the flux required for fat metabolism through synthesis and degradation of aromatic amino acids as well as precursors for acetyl-CoA. The pathways that are unique to each set are able to filter out lipid pathways responsible for BMI/obesity and dyslipidemia, as well as onset of diabetes.

Finally, our results suggest that alterations in amino acid metabolism may play an important role in adiposity and dyslipidemia in children which may be relevant to the susceptibility of metabolic diseases later in life. Our findings are consistent with recent findings which investigate the relationship between obesity in children and pathways (and their combinations) related with amino acid, lipid and glucose metabolism \citep{matsumoto2021metabolic}.

\section*{Supplementary Materials}

The Supplementary Material file: {\texttt{GPGGM\_SM.pdf}} \\ referenced throughout the manuscript is made available with this paper.

\section*{Acknowledgements}
	This research is supported by the Singapore National Research Foundation under the Translational and Clinical Research (TCR) Flagship, and Open Fund Large Collaborative Grant (OFLCG) Programmes and administered by the Singapore Ministry of Health’s National Medical Research Council (NMRC), Singapore - NMRC/TCR/004-NUS/2008; NMRC/TCR/012-NUHS/2014; OFLCG/MOH-000504. Additional funding is provided by the Singapore Institute for Clinical Sciences, Agency for Science Technology and Research (A*STAR), Singapore.

\bibliographystyle{plainnat}
\bibliography{Biblio}

\end{document}


\title{Joint modelling of association networks and longitudinal biomarkers: an application to child obesity: \\ Supplementary Materials}
	
	\author{Andrea Cremaschi $^{1,\ast}$, Maria De Iorio$^{1,2,3,4}$, Narasimhan Kothandaraman$^{1}$, \\ Fabian Yap$^{5}$, Mya Tway Tint$^{1}$, Johan Eriksson$^{1,2}$
		\\[2pt]
		\textit{$^{1}$Singapore Institute for Clinical Sciences, A*STAR, Singapore}\\
		\textit{$^{2}$Yong Loo Lin School of Medicine, National University of Singapore, Singapore}\\
		\textit{$^{3}$Division of Science, Yale-NUS College, Singapore}\\
		\textit{$^{4}$Department of Statistical Science, University College London, UK}\\
		\textit{$^{5}$Department of Paediatrics, KK Women’s and Children’s Hospital, Singapore}
		\\[2pt]
		{cremaschia@sics.a-star.edu.sg}}
	
	
	\date{}
	
	\maketitle

	
	\bigskip
	\begin{abstract}
		{This document contains the Supplementary Material for the manuscript ``Joint modelling of association networks and longitudinal biomarkers: an application to child obesity''. The document is organised as follows: Section \ref{sec:similarities-with-ppmx} shows an interesting similarity between the proposed model and existing literature on product partition models with covariates (PPMx); Sections \ref{sec:additional-figures} and \ref{sec:SM_Tables} contain additional figures and tables referenced in the main text, while Section \ref{sec:additional-results} reports additional details on the posterior inference for the GUSTO data. Section \ref{sec:SM_Alg} contains the details of the MCMC algorithm.}
		{Dirichlet Process; Gaussian process; Graph-based clustering;  Graphical models; Longitudinal data;  Metabolomics} 
	\end{abstract}

\section{Similarities with PPMx}\label{sec:similarities-with-ppmx}

We explore in this Section an interesting feature of the proposed model, linked with the existing literature on product partition models with covariates (PPMx) \citep{muller2011product}. In particular, it can be shown that the marginal distribution of the random partition induced by the DP measure can be factorised in terms including the covariates (i.e., the metabolites) within each cluster.

Let $\rho_N = \{C_1, \dots, C_{K_N}\}$ be the partition of the indices $\{1, \dots, N\}$ induced by the sample $(\bm \psi_1, \dots, \bm \psi_N)$, where we indicate by $C_j$ the $j$-th cluster of size $n_j = |C_j|$, for $j = 1, \dots, K_N$. Considering the metabolites observations $\bm M$ as covariates, and following the PPMx approach, we have that the prior for the partition $\rho_N$ is:
\begin{equation}\label{eq:rho_M_1}
	p\left(\rho_N \mid \bm M \right) = V(N, K_N, \alpha) \prod_{j = 1}^{K_N} \mathcal{C}(C_j) \mathcal{H}(\bm M^\star_j)
\end{equation}
where $\mathcal{C}(C_j)$ is the \textit{cohesion} of the $j$-th cluster $C_j$, $\mathcal{H}(\bm M^\star_j)$ is the \textit{similarity} of the metabolites for the subjects belonging to cluster $j$, denoted as $\bm M^\star_j := \{M_i : i \in C_j\}$, for $j = 1, \dots, K_N$, and $V(N, K_N, \alpha)$ is a constant derived from the marginal distribution of the partition $\rho_N$. Information about the partition is included via the cohesion function $\mathcal{C}$, while the contribution of the covariates to the clustering structure is expressed via the similarity function $\mathcal{H}$, facilitating the clustering of subjects with similar covariates. Under our modelling assumptions it can be shown that \eqref{eq:rho_M_1} is given by: 
\begin{gather}\label{eq:rho_M_2}
	p\left(\rho_N \mid \bm M \right) \propto p\left( \rho_N \right) p\left(\bm M \mid \rho_N \right) = \nonumber \\
	V(N, K_N, \alpha) \prod_{j = 1}^{K_N} \Gamma(n_j) \int \left( \prod_{i \in C_j}  p\left(\bm M_i \mid \bm \Omega_{G_j}, G_j \right) \right) P_0(d\bm \Omega_{G_j}, dG_j)
\end{gather}
The expression of $V(N, K_N, \alpha)$ can be derived from the eppf of the DP with general base measure \citep{argiento2019hierarchical, pitman2006combinatorial}. The integral can be simplified further by integrating out the precision matrix $\bm \Omega_{G_j}$, which is a-priori distributed according to a G$-$Wishart$(\nu, \bm \Psi, G)$, yielding to the ratio of normalising constants in \eqref{eq:GWishart_prior}:
\begin{gather}\label{eq:rho_M_3}
	p\left(\rho_N \mid \bm M \right) \propto p\left( \rho_N \right) p\left(\bm M \mid \rho_N \right) = \nonumber \\
	\alpha^{K_N} \frac{\Gamma(\alpha)}{\Gamma(\alpha + N)} \prod_{j = 1}^{K_N} \Gamma(n_j) \sum_{G \in \mathcal{G}_{p_M}} \frac{I_G(\nu + n_j, \bm \Psi^{(j)})}{I_G(\nu, \bm \Psi)} \pi(G)
\end{gather}
where $\bm \Psi^{(j)} = \bm \Psi + \sum_{i \in C_j} \bm M_i \bm M_i^\top$, and where the part relative to the precision matrix has been integrated out, leaving only the sum over all possible graphs of size $p_M$. The latter sum is finite and theoretically could be computed for each cluster.

\clearpage
\section{Additional Figures}\label{sec:additional-figures}

This Section includes some of the Figures discussed in the main manuscript.

\begin{figure}[ht]
	\centering
	\includegraphics[width=0.75\textwidth]{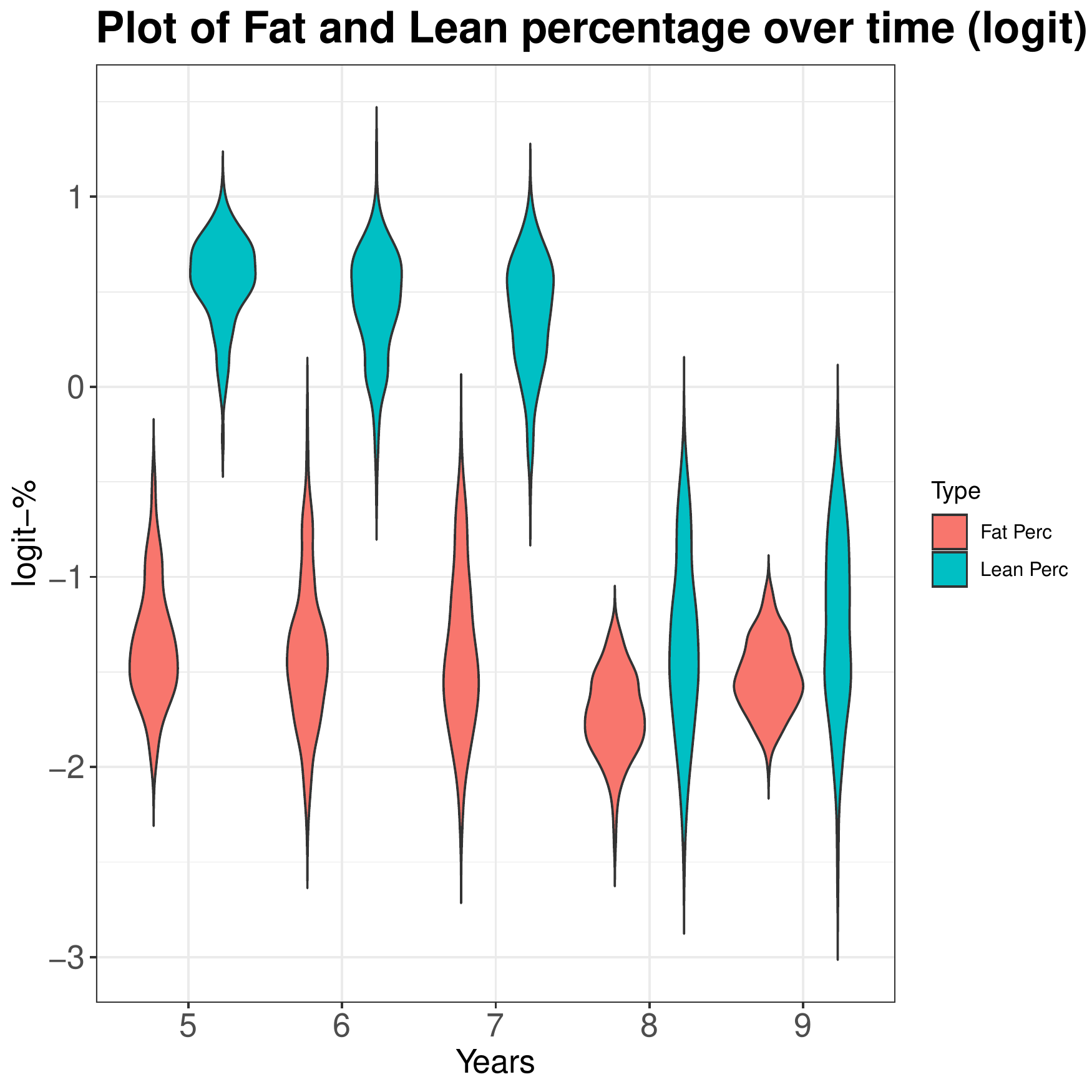}
	\caption{Longitudinal data -- Fat and lean percentages (logit transformed) over time.}
	\label{figSM:FatLean_violin}
\end{figure}

\begin{figure}[ht]
	\centering
	\includegraphics[width=0.75\textwidth]{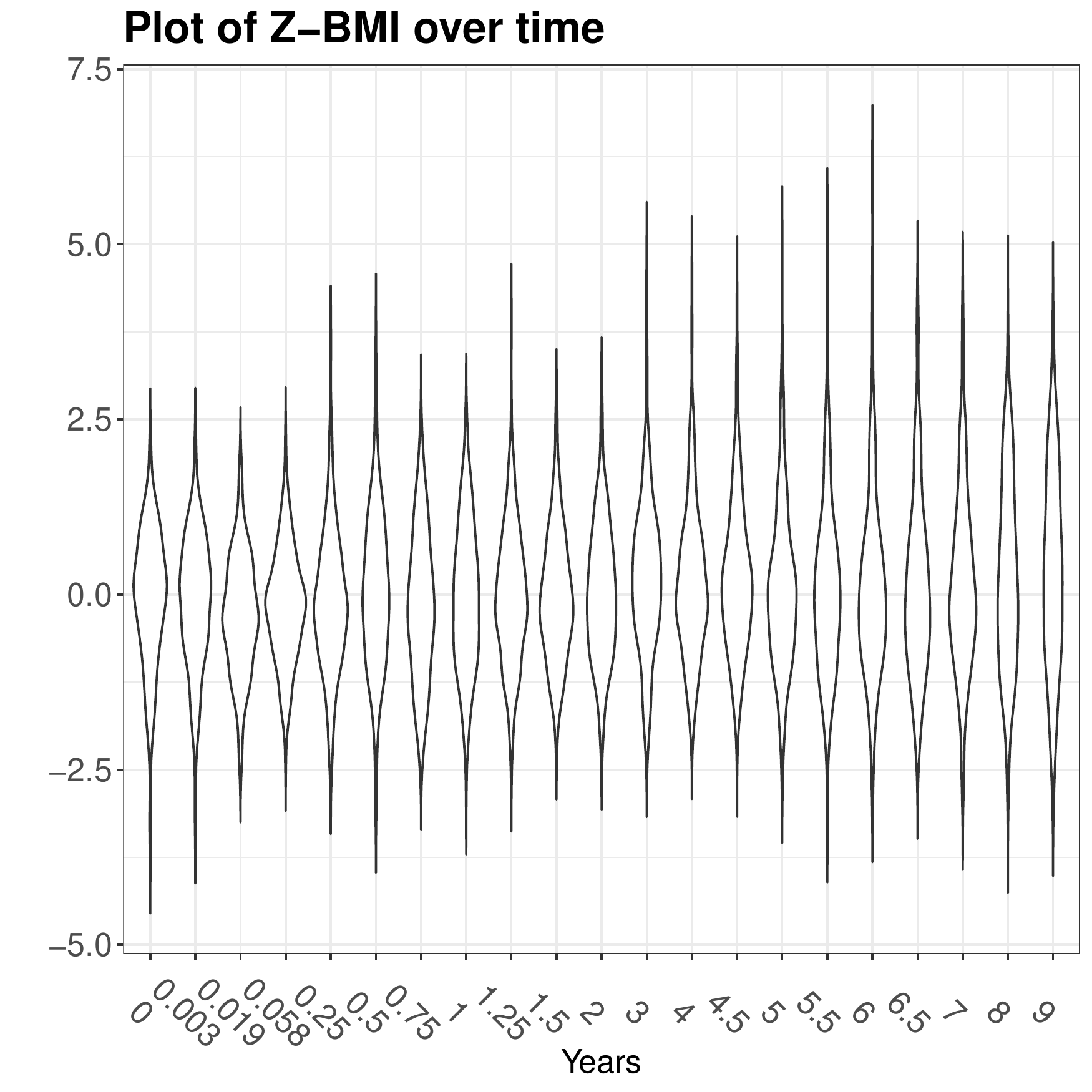}
	\caption{Longitudinal data -- Z-BMI values over time.}
	\label{figSM:ZBMI_violin}
\end{figure}

Figures \ref{figSM:FatLean_violin} and \ref{figSM:ZBMI_violin} show violin plots of the data used in the analysis, grouped by year of observation. Each Figure refer to a different set of growth indicators, namely the fat and lean percentages (logit-transformed) and the Z-BMI values.

\clearpage
\section{Additional Tables}\label{sec:SM_Tables}

Tables \ref{tab:MET_names} includes the list of metabolites used in the analysis \citep{bell2020early}.

Table \ref{tab:Covariates} includes the available covariates used in the analysis, in both the longitudinal and metabolic part of the model. The Table also includes the percentages of missing values for each covariate.


\begin{table}[!ht]
	\centering
	\caption[]{GUSTO cohort study: Metabolites used in the analysis.}
	\label{tab:MET_names}
	\begin{tabular}{c|c}
		\textbf{Metabolite name} & \textbf{Chemical family} \\\hline
		
		Clinical LDL Cholesterol & \\
		HDL Cholesterol & Cholesterol \\
		Triglycerides & \\\hline
		Phosphoglycerides & \\
		Cholines      &       Phosphoglycerides \\
		Sphingomyelins & \\\hline
		APO A1 & Apolipoproteins \\
		APO B & \\\hline
		Omega 3 & \\
		Omega 6 &     \\
		Poly-Unsaturated FA (PUFA) & \\
		Mono-Unsaturated FA (MUFA) & Fatty acids (FA) \\
		Saturated FA (SFA) & \\
		Linoleic acid & \\
		Docosahexaenoic acid (DHA)     & \\\hline       
		Alanine & \\
		Glutamine & \\
		Glycine & \\
		Histidine & \\
		Isoleucine          & Amino acids \\
		Leucine & \\
		Valine & \\
		Phenylalanine & \\
		Tyrosine & \\\hline
		Glucose              & \\
		Lactate & Glycolysis \\
		Pyruvate & \\
		Citrate & \\\hline
		$\beta$-Hydroxybutyric acid (bOHbutyrate) & \\
		Acetate              & Ketone bodies \\
		Acetoacetate & \\
		Acetone & \\\hline
		Creatinine & Fluid balance \\
		Albumin & \\\hline
		Glycoprotein acetyls & Inflammation
	\end{tabular}
\end{table}

\begin{table}[!ht]
	\caption[]{GUSTO cohort study: Time-homogeneous covariates used in the analysis.}
	\label{tab:Covariates}
	\begin{tabular}{c|c|c}
		\textbf{Variable name} & \textbf{Levels/Range} & \textbf{Missing} \\\hline
		
		mother ethnicity & \makecell{1 = Chinese, 2 = Malay, 3 = Indian} & 7.14 \% \\\cline{2-2}
		
		mother age at recruitment & $\mathbb{R}^+$ & 0.08 \% \\\cline{2-2}           
		
		mother highest education & \makecell{1 = Secondary and below, \\ 2 = Above Secondary} & 1.46 \% \\\cline{2-2}
		
		parity & \makecell{1 = Multiparous, 2 = Nulliparous} & 5.26 \% \\\cline{2-2}
		
		OGTT fasting & $\mathbb{R}^+$ & 8.74 \% \\\cline{2-2}
		
		OGTT 2h & $\mathbb{R}^+$ & 8.74 \% \\\cline{2-2}
		
		GDM WHO 1999 & {1 = No, 2 = Yes} & 8.74 \% \\\cline{2-2}    
		
		pre-pregn. BMI & $\mathbb{R}^+$ & 12.78 \% \\\cline{2-2}
		
		Gender of the baby & {1 = Male, 2 = Female} & 5.34 \% \\\cline{2-2}
		
		Gestational age & $\mathbb{R}^+$ & 5.34 \% \\
		
	\end{tabular}
\end{table}


\clearpage
\section{Additional results}\label{sec:additional-results}

\subsection{Clustering analysis}

\begin{figure}[ht]
	\subfloat[]{\includegraphics[width=0.55\textwidth]{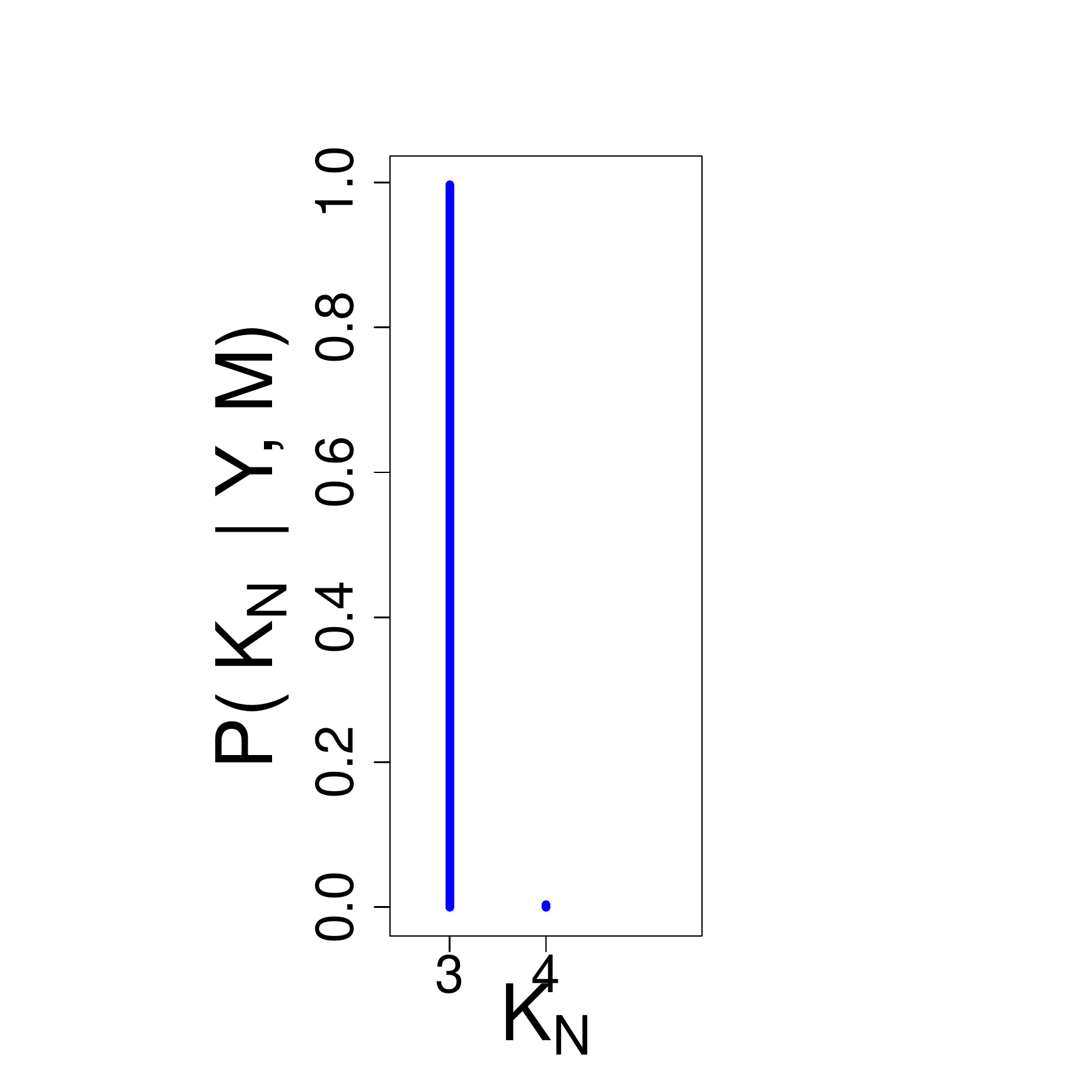}}
	\subfloat[]{\includegraphics[width=0.55\textwidth]{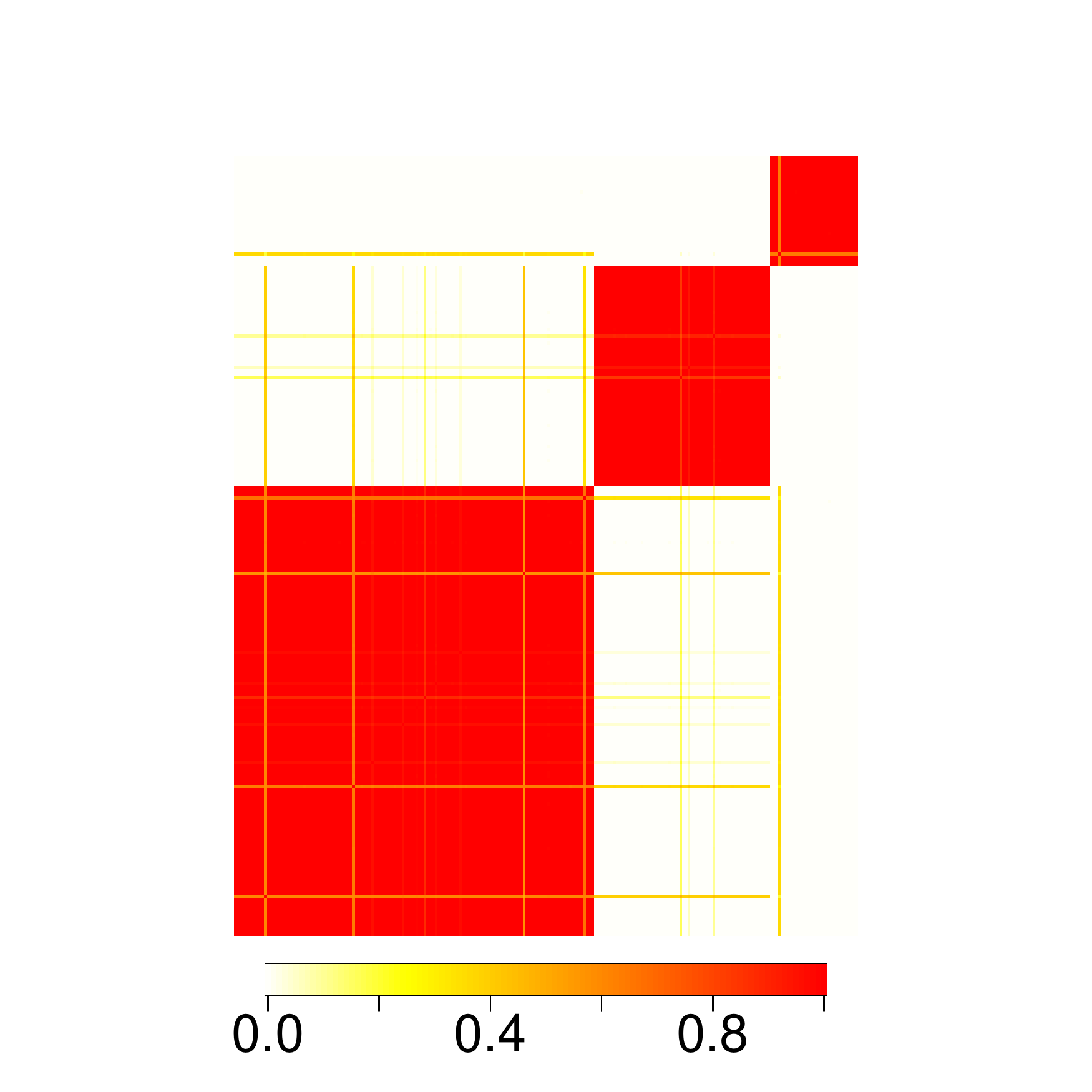}}	
	\caption{(a) Posterior distribution of the number of clusters. (b) Posterior co-clustering probabilities.}
	\label{figSM:K_N_CoClust}
\end{figure}

Figure \ref{figSM:K_N_CoClust} includes the posterior distribution of the number of clusters and the posterior co-clustering probabilities.

\begin{figure}[ht]
	\subfloat[]{\includegraphics[width=0.35\textwidth]{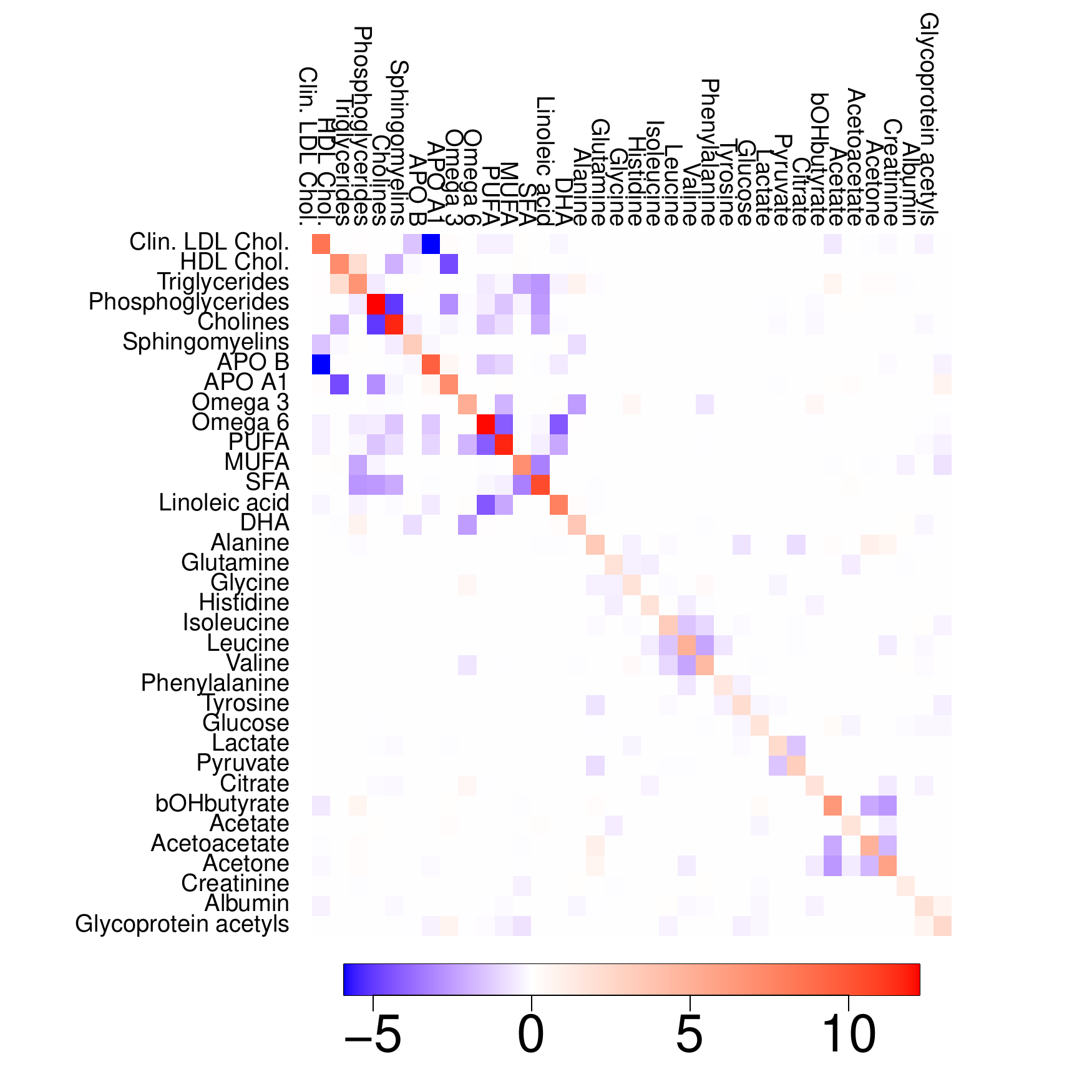}}
	\subfloat[]{\includegraphics[width=0.35\textwidth]{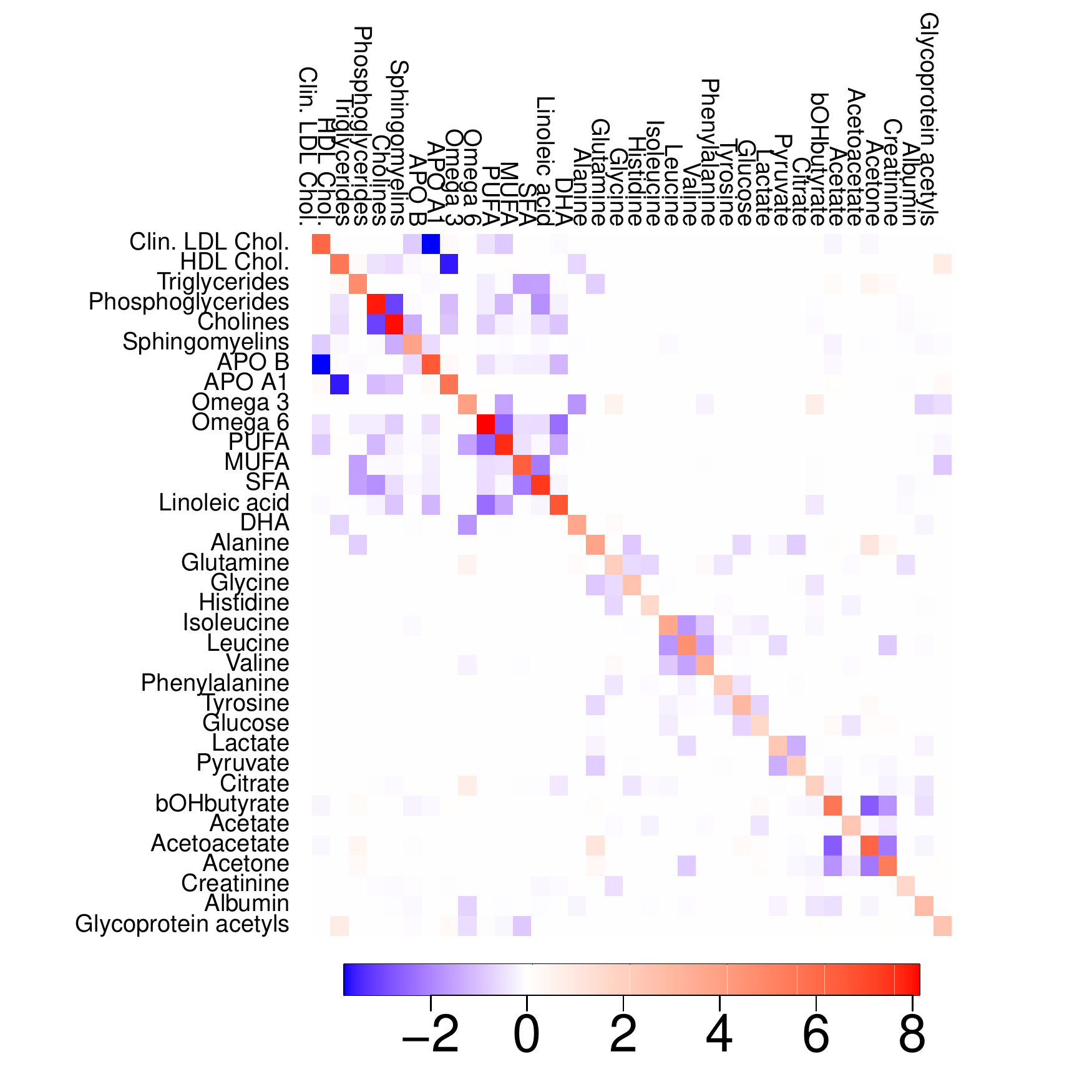}}
	\subfloat[]{\includegraphics[width=0.35\textwidth]{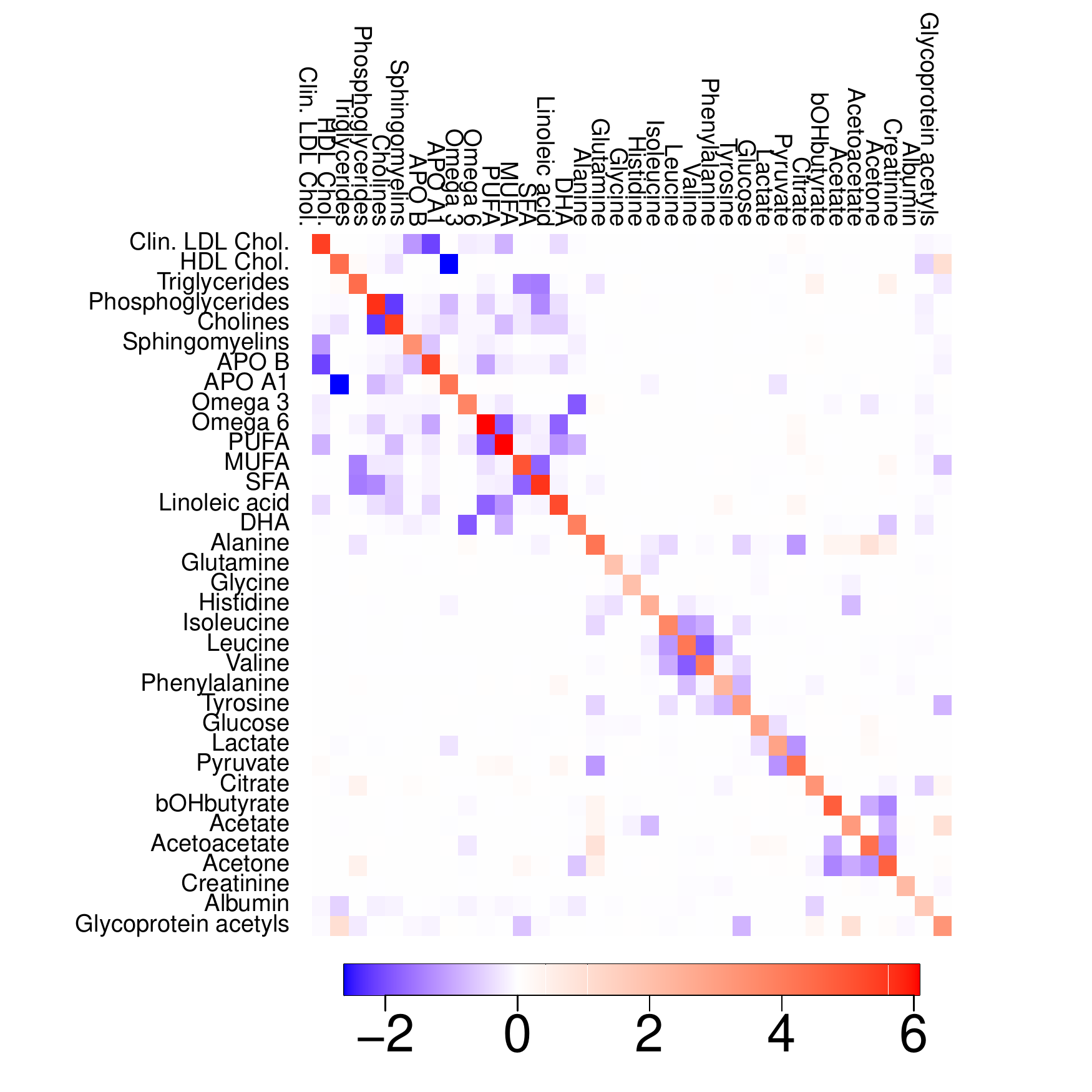}}
	\caption{Posterior mean of the precision matrices within the three identified clusters.}
	\label{figSM:OmegaG_all}
\end{figure}

Figure \ref{figSM:OmegaG_all} shows the posterior estimates of the precision matrices within each of the estimated cluster. The estimates are obtained by running an additional MCMC chain for which the partition of the subjects is fixed to the one estimated by minimising Binder's loss function.

\clearpage
\subsection{Regression on growth biomarkers}
\begin{figure}[ht]
	\centering
	\includegraphics[width=1\textwidth]{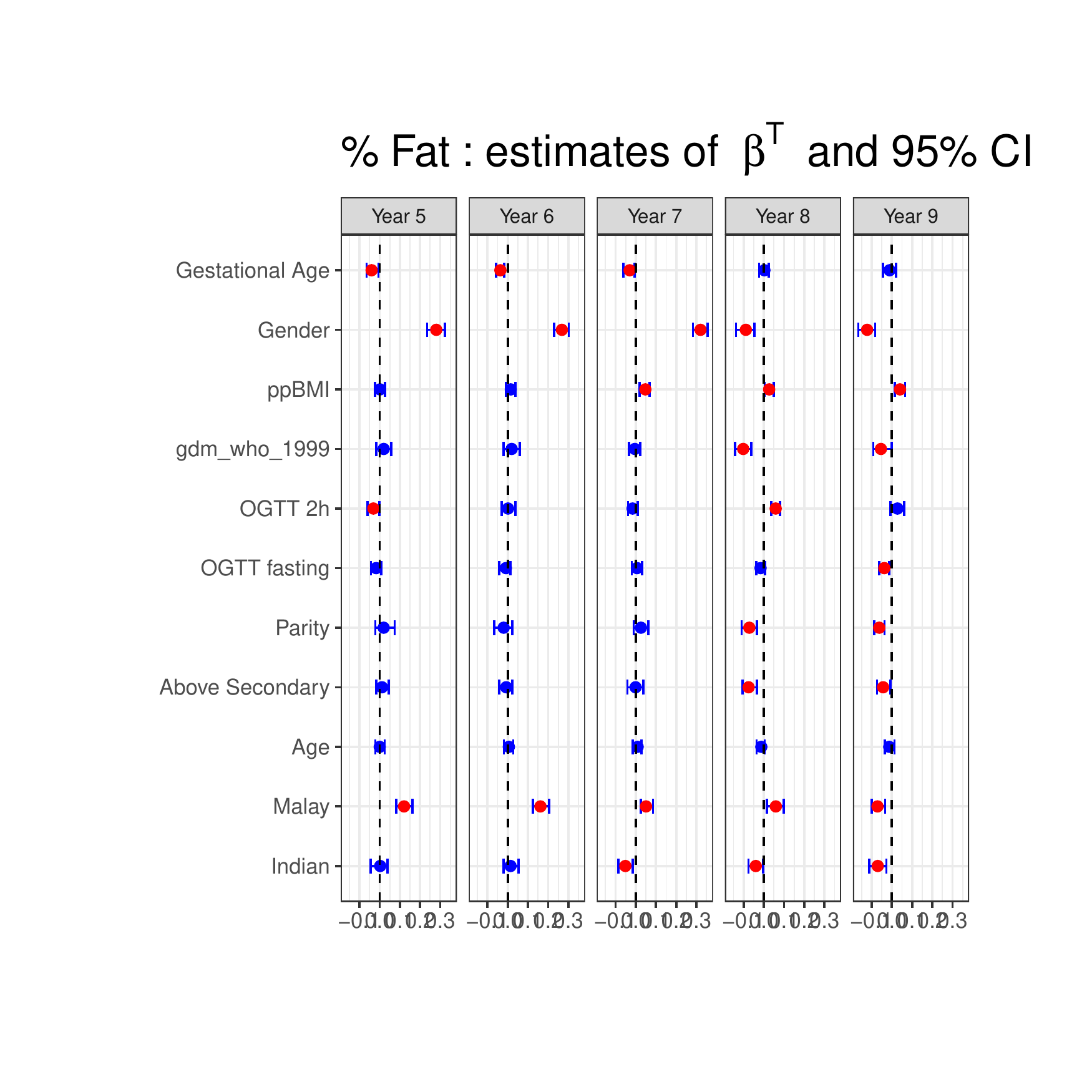}
	\caption{Posterior mean and 95\% credible intervals for the regression coefficient on the fat percentage process. Each dot represents the posterior mean of the effect of a specific covariate on a time points. The red dots indicate relevant effect of the corresponding covariate at that time.}
	\label{SMfig:beta_T_Fat}
\end{figure}

\begin{figure}[ht]
	\centering
	\includegraphics[width=1\textwidth]{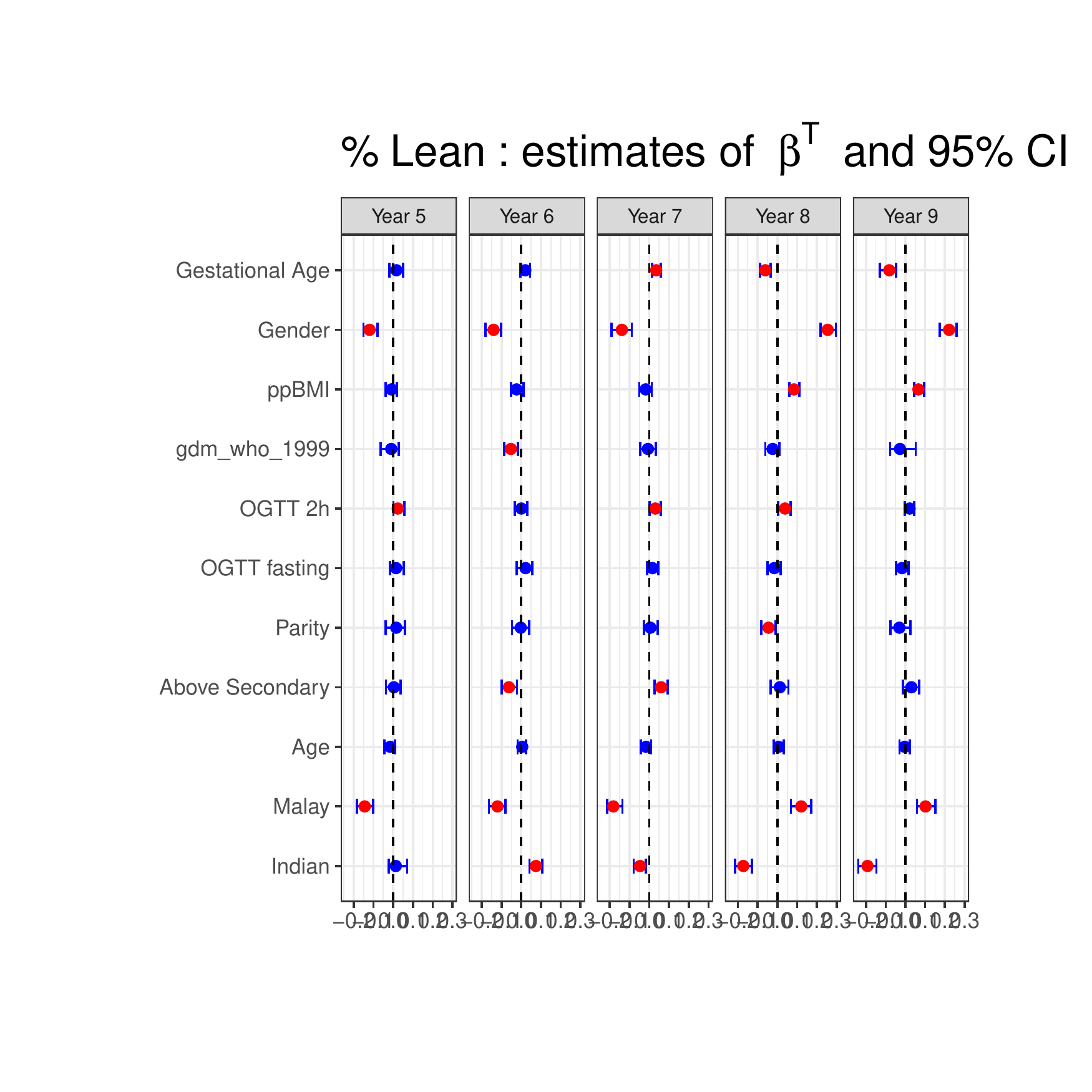}
	\caption{Posterior mean and 95\% credible intervals for the regression coefficient on the lean percentage process. Each dot represents the posterior mean of the effect of a specific covariate on a time points. The red dots indicate relevant effect of the corresponding covariate at that time.}
	\label{SMfig:beta_T_Lean}
\end{figure}

\begin{figure}[ht]
	\centering
	\includegraphics[width=1\textwidth]{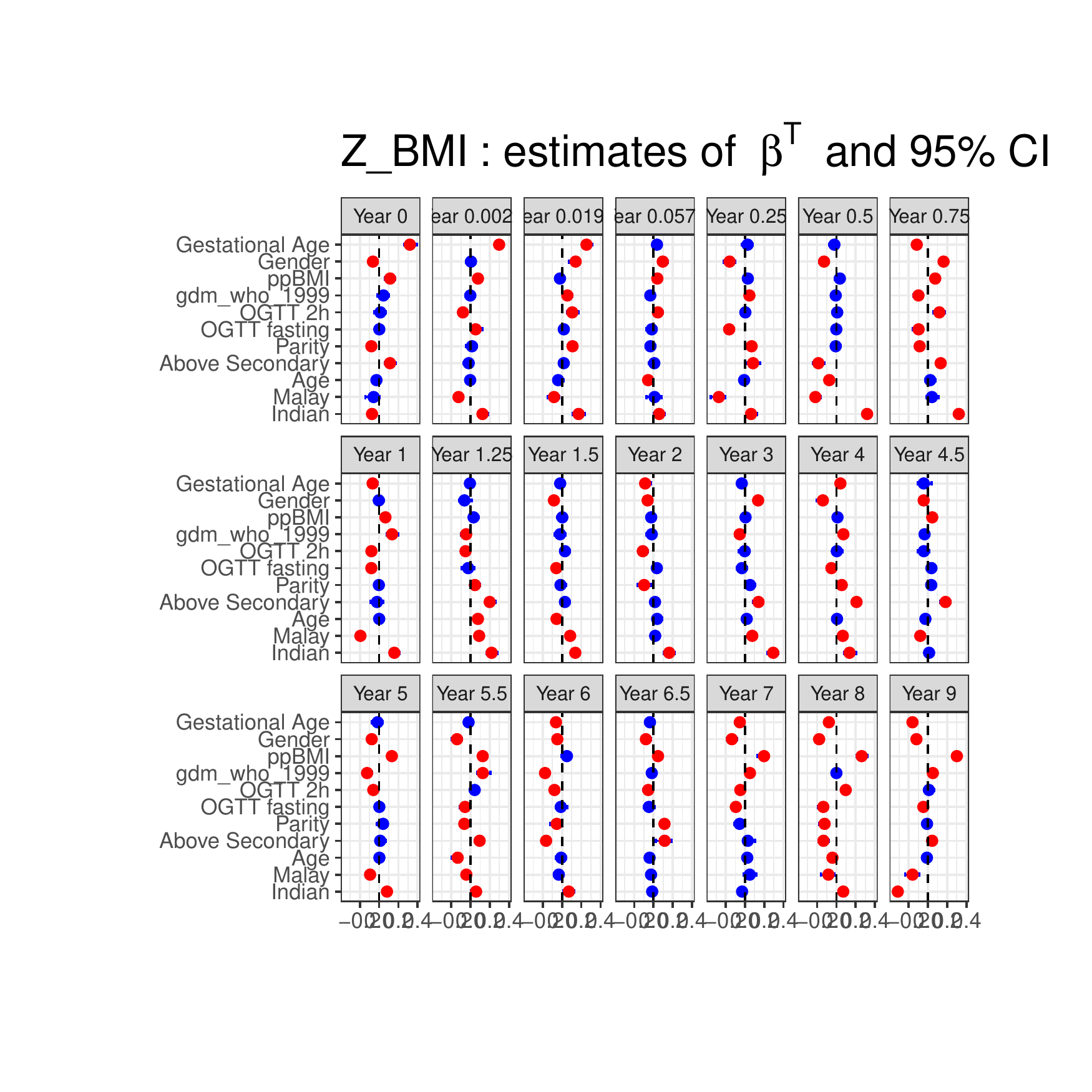}
	\caption{Posterior mean and 95\% credible intervals for the regression coefficient on the lean percentage process. Each dot represents the posterior mean of the effect of a specific covariate on a time points. The red dots indicate relevant effect of the corresponding covariate at that time.}
	\label{SMfig:beta_T_Z_BMI}
\end{figure}

We consider the effect of the covariates on the evolution of the growth indicators over time (fat and lean percentages as well as Z-BMI values). We report the posterior distribution of the corresponding regression coefficients $\bm \beta^Y$ in Figures \ref{SMfig:beta_T_Fat}, \ref{SMfig:beta_T_Lean} and \ref{SMfig:beta_T_Z_BMI}. We report the posterior 95\% CIs of the coefficients by time point, and highlight in red those coefficients whose posterior 95\% CI does not contain zero, and are therefore deemed relevant.

\subsection{Regression on Metabolite Concentrations}

\begin{figure}[ht]
	\centering
	\includegraphics[width=1\textwidth]{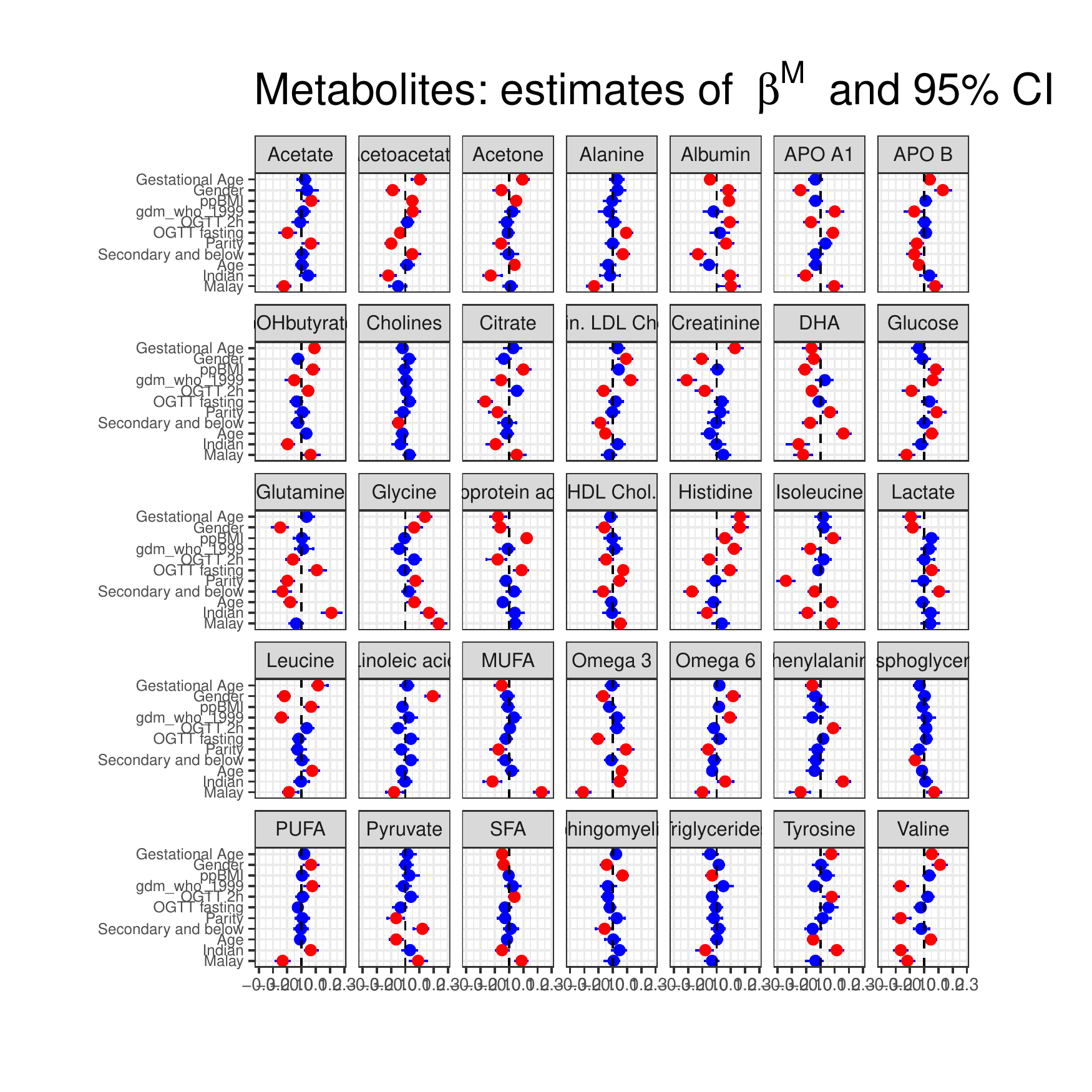}
	\caption{Posterior mean and 95\% credible intervals for the regression coefficient on the metabolites. Each dot represents the posterior mean of the effect of a specific covariate on a a metabolite. The red dots indicate relevant effect of the corresponding covariate on that metabolite.}
	\label{figSM:beta_M}
\end{figure}

We consider also the effect of the covariates on metabolite levels. We report the posterior distribution of the corresponding regression coefficients $\bm \beta^M$ in Figure \ref{figSM:beta_M}. In this case, we group the posterior 95\% CIs of the coefficients by metabolite (in total $p_M = 35$), and highlight in red those coefficients whose posterior 95\% CI does not contain zero, considered as relevant in the analysis.

\clearpage
\section{MCMC algorithm}\label{sec:SM_Alg}
In this Section, we describe the steps used to implement the MCMC algorithm. Non-conjugate parameters in the model are updated with the adaptive Metropolis-Hastings (MH) algorithm for multivariate variable \citep{haario2001adaptive}, where new candidates are proposed from a multivariate Normal distribution centred on the value of the parameter at the current iteration, and with covariance matrix equal to an appropriately re-scaled version of the sample covariance matrix obtained using the samples produced in the MCMC chain so far. The update of the DP parameters follows a P\'{o}lya Urn scheme for non-conjugate models \citep{neal2000markov, favaro2013mcmc}, adjusted for the presence of the non-conjugate graphical structure. We provide in the following the expression of the resulting acceptance probabilities.

\begin{itemize}
	\item The regression coefficients in the growth trajectories and the metabolites part of the model are assigned the following matrix-Normal prior distribution:
	$$
	p(\bm \beta) = \text{MN}_{p \times q}(\bm \beta | \bm 0, \mathbb{I}_p, \mathbb{I}_q)
	$$
	We use an adaptive MH algorithm and accept a proposed value $\bm \beta^{new}$ according to the following probability where the contribution of the proposal  distribution, being symmetric, cancels out:
	$$
	\min \left\{ 1, \frac{p(\bm \beta^{new}) \prod_{i = 1}^N \prod_{s = 1}^S \text{N}_{n_s}(\bm Y^{(s)}_i | \bm \theta^{(s)}_i + \bm \beta^{new}_s \bm X^Y_i, \mathbb{I}_{n_s}/\tau^2_s)}{p(\bm \beta^Y)\prod_{i = 1}^N \prod_{s = 1}^S \text{N}_{n_s}(\bm Y^{(s)}_i | \bm \theta^{(s)}_i + \bm \beta^Y_s \bm X^Y_i, \mathbb{I}_{n_s}/\tau^2_s)} \right\}
	$$
	for the growth trajectories and
	$$
	\min \left\{ 1, \frac{p(\bm \beta^{new}) \prod_{i = 1}^N \text{N}_p(\bm M_i \mid \bm \beta^{new} \bm X_i, \bm \Omega_{G_i})}{p(\bm \beta^M)\prod_{i = 1}^N \text{N}_p(\bm M_i \mid \bm \beta^M \bm X_i, \bm \Omega_{G_i})} \right\}
	$$
	for the metabolite concentrations.
	
	\item We assume a priori $\tau^2_s \sim inv-gamma(\tau^2_s | a_{\tau^2_s}, b_{\tau^2_s})$, and therefore:
	$$
	\tau^2_s \mid \text{rest} \sim inv-gamma \left( a_{\tau^2_s} + N  n_s/2, b_{\tau^2_s} + \frac{1}{2} \sum_{i = 1}^N\sum_{t = 1}^{n_s}(Y_{it} - \theta_{it} - \bm \beta^Y_{st} \bm X^Y_i)^2 \right)
	$$
	
	\item We assume a priori $\bm \mu_{\bm \theta} \sim \text{N}_{p_Y}(\bm \mu_{\bm \theta} | \mu_{\bm \mu_{\bm \theta}}, \bm \Sigma^{-1}_{\bm \mu_{\bm \theta}})$, and therefore:
	$$
	\bm \mu_{\bm \theta} \mid \text{rest} \sim \text{N}_{p_Y}(\bm S (\mu_{\bm \mu_{\bm \theta}} \Sigma^{-1}_{\bm \mu_{\bm \theta}} + \bm K^{-1} \sum_{j = 1}^{K_N} \bm \theta^\star), \quad \bm S = (\Sigma^{-1}_{\bm \mu_{\bm \theta}} + K_N \bm K^{-1})^{-1})
	$$	
	
	\item $\sigma^2, \phi^2, \eta^2, \xi_1, \dots, \xi_S$ are all non-conjugate and require an adaptive MH step, which includes the modification of the Gaussian covariance kernel $\bm K = \left[\bm K_{t_it_j}^{s_1s_2} \right]$, which is a function of these parameters, yielding the following acceptance probability:
		$$
	\min \left\{ 1, \frac{p(x^{new}) \prod_{j = 1}^{K_N} \text{N}_{p^Y}(\bm \theta^\star_j \mid \bm \mu_{\bm \theta}, \bm K^{new})}{p(x)\prod_{i = 1}^N \prod_{j = 1}^{K_N} \text{N}_{p^Y}(\bm \theta^\star_j \mid \bm \mu_{\bm \theta}, \bm K)} \right\}
	$$
	
	\item We propose a P\'olya urn scheme for the update of the random partition induced by the Dirichlet process (DP) prior. The centring measure $P_0$ is non-conjugate in the proposed model, thus requiring the use of suitable algorithms \citep{neal2000markov, favaro2013mcmc}. Notice that the support of the centring measure $P_0$ is mixed (i.e., contains atoms represented by the graphs, but also a diffuse part for the mean vector of the growth trajectories). Thanks to the theoretical properties of the Dirichlet process discussed in Section 2 of the main manuscript, the steps of the algorithm are analogous of the case of a diffuse centring measure. An important difference, however, is that we expect ties in the values of the parameters associated with the clusters, $\bm \psi^\star_1, \dots, \bm \psi^\star_{K_N}$, specifically in the values of the graphs.
	
	\item Given the partition of the subjects, the unique values of the mean vectors of the growth trajectories $\bm \theta^\star_1, \dots, \bm \theta^\star_{K_N}$ have a conjugate full-conditionals:
	$$
	\bm \theta^\star_j \mid \text{rest} \sim \text{N}_{p_Y}(\bm S (\bm K^{-1}\mu_{\bm \theta} + diag(1 / \bm \tau^2) \sum_{i \in C_j} (\bm Y_i - \bm \beta^Y \bm X_i) ), \bm S = (\bm K^{-1} + diag(n_j / \bm \tau^2))^{-1})
	$$
	where $n_j = \mid C_j \mid$ for $j = 1, \dots, K_N$ and $\bm \tau^2$ is a vector composed of the process-specific variances $\tau^2_s$ replicated $n_s$ times, for $s = 1, \dots, S$.
	
	\item The unique values in the graph structure part of the DP are updated conditionally to the subjects within each cluster, using the \texttt{BDgraph} \citep{mohammadi2015bayesian}.

	\item Missing values in the growth trajectories $\bm Y^{(s)}$, for $s = 1, \dots, S$ or in the metabolite concentrations $\bm M$ are updated by sampling them from their full conditional distribution, which is readily obtainable through results on the conditional distribution of a multivariate Gaussian.
\end{itemize}

\bibliographystyle{plainnat}
\bibliography{Biblio}